\documentclass[aip,pof,amsmath,amssymb,reprint]{revtex4-1}

\usepackage{tabularx}
\newcolumntype{C}[1]{>{\centering\arraybackslash}p{#1}}
\usepackage{units}
\usepackage{dcolumn}
\usepackage{siunitx}
\usepackage{color}
\usepackage[capitalise]{cleveref}
\usepackage{tikz}
\usepackage{pgfplots}
\usepgfplotslibrary{groupplots}
\usetikzlibrary{spy,decorations.fractals,shapes.misc}
\usetikzlibrary{decorations.markings}
\usepgfplotslibrary{colormaps}
\usepgfplotslibrary{colorbrewer}
\pgfplotsset{cycle list/Dark2}

\usepgfplotslibrary{external}
\tikzset{external/system call={pdflatex \tikzexternalcheckshellescape -halt-on-error
-interaction=batchmode -jobname "\image" "\texsource" && 
pdftops -eps "\image.pdf"}}

\usepackage[caption=false]{subfig}
\usepackage{epstopdf}

\usepackage{booktabs,multirow}

\begin{document}

\title[Multi-species modeling in the ESBGK method for monatomic species]{Multi-species modeling in the particle-based ESBGK method for monatomic gas species}

\author{M. Pfeiffer}
 \email{mpfeiffer@irs.uni-stuttgart.de}
\affiliation{%
Institute of Space Systems, University of Stuttgart, Pfaffenwaldring 29, D-70569
   Stuttgart, Germany
}%

\author{A. Mirza}
 \email{mirza@boltzplatz.eu}
\author{P. Nizenkov}
 \email{nizenkov@boltzplatz.eu}
\affiliation{%
boltzplatz - Numerical Plasma Dynamics, Technology Transfer Initiative GmbH at the University of Stuttgart, Nobelstraße 15, 70569 Stuttgart, Germany
}%

\date{\today}

\begin{abstract}
Multi-species modeling is implemented for the particle-based ellipsoidal statistical Bhatnagar-Gross-
Krook (ESBGK) for monatomic species in the open-source plasma simulation suite PICLas. After a literature review on available multi-species extensions of the kinetic model equations and approaches for the determination of the transport coefficients, Brull's model is implemented for the former and Wilke's mixing rules and collision integrals for the latter. The implementation is verified with three simulation test cases: a simple reservoir, a supersonic Couette flow and the hypersonic flow around a 70$^{\circ}$ blunted cone. The simulation results are compared with the Direct Simulation Monte Carlo (DSMC) method, where good overall agreement can be achieved. However, the determination of the transport coefficients through collision integrals offers better agreement with the DSMC results at acceptable computational cost. For the last test case, a comparison of the computational duration is presented.
\end{abstract}


\maketitle
\section{Introduction}

Numerical simulations of fluid dynamics in space applications such as atmospheric entry maneuvers and in-space propulsion systems pose considerable challenges for the applied numerical methods. Large density gradients spanning from the continuum regime to the free molecular flow and the consequent breakdown of the continuum assumption require the utilization of numerical methods with an extended applicability. A well-established method for the simulation of rarefied gas flows is the Direct Simulation Monte Carlo (DSMC) method \cite{Bird1994}. Although applicable in the continuum regime, the microscopic treatment of the gas flow becomes computationally infeasible. Thus, a coupling of DSMC with a computationally efficient method is desirable. While conventional computational fluid dynamics (CFD) approaches based on the solution of the Navier-Stokes equations offer a fast numerical solution in the continuum regime, the bidirectional coupling with the DSMC method is cumbersome due to the utterly different approaches of the methods \cite{zhang2019particle}. Recently, particle-based continuum methods gained traction as an alternative solution, bridging the gap in the transitional regime. A recent review of several such methods was conducted by \citet{Pfeiffer2019a}.

One of the most promising approaches is the Bhatnagar-Gross-Krook model \cite{bhatnagar1954model}, which is an approximation of the collision integral of the Boltzmann equation. The assumption is that the distribution function relaxes towards a target distribution function, which can have different forms. While the original BGK model using the Maxwellian distribution cannot reproduce the correct Prandtl numbers, different alternative distribution functions exist. Two popular functions are used for the simulation of rarefied gas flows in high-speed applications: the Ellipsoidal Statistical (ESGBK)\cite{Holway1966} and Shakhov (SBGK)\cite{Shakhov1968} models. In the continuum limit, both methods can be used to derive the Navier-Stokes equations. While both models have been implemented in PICLas\footnote{PICLas is a flexible particle-based plasma simulation suite. Available online at https://github.com/piclas-framework/piclas.}$^{,}$\cite{Pfeiffer2018a,Pfeiffer2018b} and extensively tested for
atmospheric re-entry \cite{Pfeiffer2019c} and nozzle expansion \cite{Pfeiffer2019a} flows in the particle-based context, this has been done for single component gases. The focus is this paper is the extension of the particle-based ESBGK model to gas mixtures. For this purpose, two major challenges have to be addressed.

\subsection{Extension of the kinetic model equations}

First, a proper mathematical model is required which fulfills the consistency requirements of the Boltzmann collision operator such as conservation, equilibrium, H-theorem and positivity of the density and temperature fields. Moreover, it has to satisfy the indifferentiability principle \cite{Andries2002}, meaning that in case of identical species in terms of masses and cross-sections, the total distribution function reduces to the single species system. Finally, these models have to be able to correctly reproduce the transport coefficients (diffusion by Fick's law, viscous stress by Newton's law, and thermal conductivity by Fourier's law) in the continuum limit. Available models can be categorized based on the treatment of the collision term \cite{Klingenberg2018a}: single- and multi-relaxation modeling, where the difference is whether a single or multiple relaxation rates are employed for self- and cross-collisions. On the one hand, single-relaxation models are computationally more efficient and less complex than multi-relaxation models. On the other hand, multi-relaxation models with different relaxation rates are likely to be better suited for mixtures, where the species differ substantially. Since many of the models were derived mathematically but not applied to a realistic application, it is not clear whether the more complex multi-relaxation model are absolutely required for most of the application cases. Therefore, the focus of this project will be on single-relaxation models. However, \citet{Klingenberg2018b} are developing a potential multi-relaxation model, which might be considered in future work.

Single-relaxation models for the ESBGK model, which satisfy the indifferentiability principle, were introduced by \citet{Groppi2011} and \citet{Brull2014}, extending the ideas introduced by \citet{Andries2002} for the original BGK model. The model by \citet{Groppi2011} is able to reproduce the correct diffusion coefficient and viscous stress and the model by \citet{Brull2014} is able to reproduce the correct viscous stress and thermal conductivity. To rectify the shortcomings, \citet{Todorova2019} introduced an extension to the model by \citet{Groppi2011} to reproduce the correct thermal conductivity in the continuum limit by adding an additional free parameter. The focus of this paper is Brull's model with two different approaches for the determination of the transport coefficients.

\subsection{Determination of the transport coefficients for mixtures}

Second, a method for the determination of transport coefficients of the gas mixture has to be devised. Here, much of the work done for conventional computational fluid dynamics can be capitalized on. One of the most widely used approaches are the mixing rules by \citet{Wilke1950}, which were derived from kinetic gas theory. \citet{Palmer2003} gives an overview of available mixing rules and approximative methods to determine transport coefficients, including a comparison of Wilke's mixing rules with two more advanced models. A more recent approach with fewer assumptions is to calculate the transport coefficients directly from collision integrals as outlined by \citet{Hirschfelder1964}. For this purpose, different fits for the collision integrals can be utilized to speed up calculation times (e.g. by \citet{Kestin1984}, \citet{Capitelli2000}, and \citet{Wright2005,Wright2007a}, who give an overview over the available collision integral data for the atmospheres of Earth, Mars, and Venus). Most of these collision integrals assume an attractive-repulsive intermolecular potential. However, to allow for a comparison with the Direct Simulation Monte Carlo method in PICLas, either collision integrals using the Variable Hard Sphere (VHS) have to be implemented as given by \citet{Stephani2012} or a collision model using the Lennard-Jones potential has to be implemented in the DSMC method as presented by \citet{Venkattraman2012}. The former approach is chosen for this paper.

\section{Theory}
The Boltzmann equation describes the gas kinetic behavior of the particle distribution function 
$f_s=f(\mathbf x, \mathbf v, t)$ at position $\mathbf x$ and velocity $\mathbf v$ of the species $s$
\begin{equation}
\frac{\partial f_s}{\partial t} + \mathbf v \frac{\partial f_s}{\partial \mathbf x} = \left.\frac{\delta f_s}{\delta t}\right|_{\mathrm{coll}},
\end{equation}
where external forces are neglected. Furthermore, $\left.\delta f_s/\delta t\right|_{\mathrm{coll}}$ is the 
collision term for a mixture of $M$ gases, which can be described by the Boltzmann collision integral
\begin{eqnarray}
&&\left.\frac{\partial f_s}{\partial t}\right|_{\mathrm{coll}}= \\ 
&&\sum_{k=1}^M \int_{\mathbb{R}^3}\int_{S^2}\mathcal B_{s,k}
\left[f_s(\mathbf v')f_k(\mathbf v_*')-f_s(\mathbf v)f_k(\mathbf v_*)\right]d\mathbf n d\mathbf v_* \nonumber.
\end{eqnarray}
Here, $S^2\subset\mathbb{R}^3$ is the unit sphere, $\mathbf n$ is the unit vector of the scattered velocities, $\mathcal B_{s,k}$ is the collision
kernel and the superscript $'$ denotes the post-collision velocities. The multiple integration of this collision term makes is difficult to compute. 
The macroscopic flow values particle density $n$, flow velocity $u$ and temperature $T$ of each species $s$ are defined as:
\begin{alignat}{1}
&n_s=\int_{\mathbb{R}^3} f_s\,d\mathbf v, \quad n_s \mathbf u_s =\int_{\mathbb{R}^3} \mathbf v f_s\,d\mathbf v, \\
&\mathcal{E}_s=\frac{3}{2}k_{\mathrm{B}} T_s = \frac{m_s}{2n_s}\int_{\mathbb{R}^3} \mathbf c_s^2 f_s\,d\mathbf v,\\
&E_s=\frac{1}{2}m_sn_s\mathbf u_s^2 + n_s\mathcal{E}_s,
\end{alignat}
with the thermal particle velocity $\mathbf c_s=\mathbf v-\mathbf u_s$. Furthermore, the macroscopic mean values of the flow are given by:
\begin{alignat}{1}
&n=\sum_{k=1}^M n_k,\quad \rho=\sum_{k=1}^M m_k n_k,\quad \rho \mathbf u = \sum_{k=1}^M m_k n_k \mathbf u_k,\\
&n\mathcal{E} + \frac{\rho}{2}\mathbf u^2 = E=\sum_{k=1}^M E_k,
\quad \mathcal{E}=\frac{3}{2}k_{\mathrm{B}} T.
\end{alignat}

\subsection{ESBGK Mixture Model}

The ESBGK mixture model of Brull~\cite{Brull2014} approximates the collision term using one relaxation term per species, where the distribution function relaxes towards a target distribution function $f_s^{\mathrm{ES}}$ with a 
certain relaxation frequency $\nu$:
\begin{equation}
\left.\frac{\partial f_s}{\partial t}\right|_{\mathrm{coll}}=\nu\left(f_s^{\mathrm{ES}}-f_s\right).
\label{eq:bgkmain}
\end{equation}
The target velocity distribution function $f_s^{\mathrm{ES}}$ is given by 
\begin{equation}
f_s^{\mathrm{ES}}=\frac{n_s}{\sqrt{\det \mathcal A}} \left(\frac{m_s}{2\pi k_{\mathrm{B}} T}\right)^{3/2} \exp\left[-\frac{m_s\mathbf c^T \mathcal A^{-1} \mathbf c}{2k_{\mathrm{B}} T}\right]
\label{eq:esbgkdist}
\end{equation}  
with the anisotropic matrix 
\begin{equation}
\mathcal A = \mathcal I - \frac{1-\alpha Pr}{\alpha Pr}\left(\frac{\mathcal P}{k_{\mathrm{B}}T/m}-\mathcal I\right).
\label{eq:asintropmat}
\end{equation}
The anisotropic matrix $\mathcal A$ consists of the identity matrix $\mathcal I$ and the pressure tensor $\mathcal P$,
\begin{equation}
\mathcal P =\frac{1}{\rho}\sum_{k=1}^M m_k\int (\mathbf v-\mathbf u) (\mathbf v-\mathbf u)^T f_k\,d\mathbf v,
\end{equation}
which are both symmetric. 
Additionally, $\mathbf c=\mathbf v -\mathbf u$ is the thermal particle velocity determined from the 
particle velocity $\mathbf v$ and the average flow velocity $\mathbf u$. $Pr$ is the targeted Prandtl number of the gas mixture and $\alpha$ is a mass fraction and density fraction dependent variable of the model
\begin{equation}
  \alpha = \sum_{k=1}^M \frac{n_k}{n}\frac{m}{m_k},\quad m=\sum_{k=1}^M \frac{n_k}{n}m_k.
\end{equation}
The relaxation frequency $\nu$ of the model is defined by 
\begin{equation}
\nu = \frac{nk_{\mathrm{B}} T}{\mu}\alpha Pr
\end{equation}
with the viscosity of the mixture $\mu$. The ESBGK model of Brull~\cite{Brull2014} produces a positive definite matrix $\mathcal A$ for Prandtl numbers in the range of 
$\left[\frac{2}{3\alpha},\infty\right[$, which depends on the involved species masses and the mole fractions.
More precisely, this condition is too restrictive according to \citet{mathiaud2016fokker}. It is demonstrated that the matrix is positive definite as long as the Prandtl number used in the scheme $Pr^*$ 
is chosen as
\begin{eqnarray}
\alpha Pr^*  =& \frac{1}{1-\tilde\nu}, \\
\tilde\nu  =& \max\left(1-\frac{1}{\alpha Pr},-\frac{k_{\mathrm{B}}T/m}{\lambda_{max}-k_{\mathrm{B}}T/m}\right),
\end{eqnarray}
with $\lambda_{max}$ being the maximum eigenvalue of $\mathcal P$. Nevertheless, in every simulation of this paper the case never occurred that the target Prandtl number could not be reached.
The Brull-ESBGK model for mixtures reproduces the Maxwellian distribution in the equilibrium state and fulfills the H-theorem. Furthermore, it fulfills the indifferentiability principle which means that the model reduces to a single species model for identical species~\cite{Brull2014}. 

\subsection{Gas mixture properties}
For the calculation of the gas mixture viscosity assuming a variable hard sphere (VHS) interaction potential, two different approaches are tested: the approximation of the mixture properties using Wilke's mixture rules\cite{Wilke1950} and the first approximation of the transport properties using collision integrals\cite{Hirschfelder1964}.

\subsubsection{Wilke's mixing rules}
For this approach, the well known exponential ansatz of the viscosity $\mu_k$
\begin{equation}
\mu_k=\mu_{\mathrm{ref},k}\left(\frac{T_k}{T_{\mathrm{ref},k}}\right)^{\omega_{\mathrm{VHS},k}}
\end{equation}
is used for each species $k$. Here, $T_{\mathrm{ref},}$ is a reference temperature, $\mu_{\mathrm{ref},k}$ the reference dynamic viscosity at $T_{\mathrm{ref},k}$ \cite{burt2006evaluation}
and $\omega_{\mathrm{VHS},k}$ is a parameter of the VHS model. For a VHS gas the reference dynamic viscosity can be calculated with the VHS reference diameter $d_{\mathrm{ref},k}$:
\begin{equation}
\mu_{\mathrm{ref},k}=\frac{30\sqrt{m_kk_{\mathrm{B}}T_{\mathrm{ref},k}}}{4\sqrt{\pi}(5-2\omega_{\mathrm{VHS},k})(7-2\omega_{\mathrm{VHS},k})d_{\mathrm{ref},k}^2}.
\end{equation}
The mixture viscosity is calculated using Wilke's mixture rule\cite{Wilke1950}:
\begin{equation}
\mu=\sum_{k=1}^M n_k \frac{\mu_k}{\Phi_k},\quad \Phi_k=\sum_{r=1}^M n_r\frac{\left(1+\sqrt{\frac{\mu_k}{\mu_r}}\left(\frac{m_r}{m_k}\right)^{1/4}\right)^2}
    {\sqrt{8\left(1+\frac{m_k}{m_r}\right)}}.
\end{equation}

The Prandtl number of the gas mixture is defined as 
\begin{equation} 
Pr = c_p\frac{\mu}{K}
\end{equation}
with the thermal conductivity $K$ of the mixture and the specific heat $c_p=\frac{5k_{\mathrm{B}}}{2m}$ evaluated with the mixture mass $m$.
The thermal conductivity of each species $K_k$ is calculated using the Eucken's relation with the viscosity\cite{Palmer2003}:
\begin{equation}
K_k=\frac{15}{4}\frac{\mu_k k_{\mathrm{B}}}{m_k}.
\end{equation}
Afterwards, the thermal conductivity of the mixture $K$ is again calculated using Wilke's mixture rule:
\begin{equation}
K=\sum_{k=1}^M n_k \frac{K_k}{\Phi_k}.
\end{equation}

\subsubsection{First approximation of transport properties}
The first approximation to the viscosity of species $k$ depending on the collision integral $\Omega_k^{(2)}(2)$ is given by\cite{Hirschfelder1964}
\begin{equation}
\mu_k=\frac{5k_{\mathrm{B}}T}{8\Omega_k^{(2)}(2)}.
\end{equation}
The mixture viscosity is determined by 
\begin{equation}
\mu=\sum_{k=1}^M b_k,
\end{equation}
where $b_k$ is the contribution of each species to the total mixture viscosity and is determined by solving the system
\begin{eqnarray}
\chi_k &= b_k\left(\frac{\chi_k}{\mu_k} + \sum_{r\neq k}\frac{3\chi_r}{(\rho_r'+\rho_k')D_{kr}}\left(\frac{2}{3}+\frac{m_r}{m_k}A_{kr}\right)\right)\nonumber\\
    &-\chi_k\sum_{r\neq k}\frac{3 b_r}{(\rho_r'+\rho_k')D_{kr}}\left(\frac{2}{3}-A_{kr}\right)
\end{eqnarray}
with the mole fraction $\chi$, the density $\rho_r'$ of species $r$ when pure at pressure and temperature of the actual gas mixture, the parameter $A_{kr}$ defined by
\begin{equation}
A_{kr}=\frac{\Omega_{kr}^{(2)}(2)}{5\Omega_{kr}^{(1)}(1)}
\end{equation}
and the binary diffusion coefficient with the reduced mass $m^*_{kr}$:
\begin{equation}
D_{kr}=\frac{3k_{\mathrm{B}} T}{16nm^*_{kr}\Omega_{kr}^{(1)}(1)}.
\end{equation}
The mixture thermal conductivity $K$ is calculated by
\begin{equation}
K=\sum_{k=1}^M a_k,
\end{equation}
with $a_k$ being the species contribution to the total mixture thermal conductivity. The factors $a_k$ are determined by solving the system
\begin{eqnarray}
\chi_k &= a_k\left[\frac{\chi_k}{K_k} + \sum_{r\neq k}\frac{\chi_r}{5k_{\mathrm{B}} n D_{kr}} \right.\\
  &\times\left(6\left(\frac{m_k}{m_r + m_k}\right)^2- (5-4B_{kr})\left(\frac{m_r}{m_r + m_k}\right)^2 \right.\nonumber\\
  &\left.\left. +8\frac{m_rm_k}{(m_k+m_r)^2}A_{kr}\right)\right]\nonumber\\
  &-\chi_k\sum_{r\neq k}a_r\frac{m_rm_k}{(m_k+m_r)^2}(5k_bn D_{kr})^{-1}\nonumber \\
  &\times\left(11-4B_{kr}-8A_{kr}\right).\nonumber
\end{eqnarray}
Here, $K_k$ is the first approximation of the thermal conductivity of species $k$
\begin{equation}
K_k=\frac{25c_V k_{\mathrm{B}} T}{16 \Omega_k^{(2)}(2)},\quad c_V=\frac{3k_b}{2m_k}
\end{equation}
and the parameter $B_{kr}$ is defined by
\begin{equation}
B_{kr}=\frac{5\Omega_{kr}^{(1)}(2)-\Omega_{kr}^{(1)}(3)}{5\Omega_{kr}^{(1)}(1)}
\end{equation}
The collision integrals for the VHS model are given by \citet{Stephani2012}:
\begin{eqnarray}
\Omega_{kr}^{\mathrm{VHS},(1)}(1)&=\frac{\pi}{2}d_{\mathrm{ref}}^2\sqrt{\frac{k_{\mathrm{B}}T}{2\pi m^*_{kr}}}\left(\frac{T_{\mathrm{ref}}}{T}\right)^{\omega-1/2}\frac{\Gamma(7/2-\omega)}{\Gamma(5/2-\omega)}\nonumber\\
\Omega_{kr}^{\mathrm{VHS},(2)}(2)&=\frac{\pi}{3}d_{\mathrm{ref}}^2\sqrt{\frac{k_{\mathrm{B}}T}{2\pi m^*_{kr}}}\left(\frac{T_{\mathrm{ref}}}{T}\right)^{\omega-1/2}\frac{\Gamma(9/2-\omega)}{\Gamma(5/2-\omega)}\nonumber\\
B^{\mathrm{VHS}}_{kr}&=\frac{5\Gamma(9/2-\omega)-\Gamma(11/2-\omega)}{5\Gamma(7/2-\omega)},
\end{eqnarray}
with the VHS parameters $d_{\mathrm{ref}}$, $T_{\mathrm{ref}}$ and $\omega$.

\section{Implementation}
The proposed ESBGK particle method for mixtures is implemented in the PIC-DSMC-BGK code PICLas \cite[]{Munz2014,fasoulas2019combining} as described in detail by Pfeiffer\cite{Pfeiffer2018a, Pfeiffer2018b}.
The main concept of the particle ESBGK method, especially the energy and momentum conservation is based on the works of \citet{gallis2011investigation,gallis2000application}.
The ESBGK particle method has many similarities to the DSMC method: particles are moved on a simulation mesh,
collide with boundaries and the microscopic particle properties are sampled to calculate macroscopic values in the same manner. But instead of performing binary collisions between particles, each particle of species $s$ in a cell relaxes with the probability
\begin{equation}
P=1-\exp\left[-\nu \Delta t\right]
\label{eq:bgkrelax}
\end{equation}
according to Eq. \eqref{eq:bgkmain} towards the target distribution $f_s^{\mathrm{ES}}$. For this, the relaxation frequency $\nu$ is evaluated in each time step for each cell, depending on the targeted Prandtl number and the mixture viscosity in the cell.

A particle chosen to relax receives a new particle velocity sampled from the target distribution with the corresponding particle mass of the species. 
The detailed description of the sampling process for different target distributions (e.g. ESBGK or SBGK) can be found in \citet{Pfeiffer2018a}.
As proposed by \citet{gallis2011investigation}, a symmetric transformation matrix $\mathcal S$ can be defined to describe the anisotropic matrix $\mathcal A$ from Eq.~\eqref{eq:asintropmat}: $\mathcal A = \mathcal S \mathcal S$.
Furthermore, a normalized thermal velocity vector $\mathbf C$ is defined as such that $\mathbf c= \mathcal S \mathbf C$. 
Using these definitions, the argument of the exponential function in Eq. \eqref{eq:esbgkdist} becomes
\begin{equation}
\mathbf c^T \mathcal A^{-1} \mathbf c = (\mathcal S\mathbf C)^T \mathcal S^{-1} \mathcal S^{-1} \mathcal S\mathbf C = \mathbf C^T \mathbf C  
\label{eq:smat}
\end{equation}
using $(\mathcal S\mathbf C)^T=\mathbf C^T \mathcal S^T=\mathbf C^T \mathcal S$ due to the fact that $\mathcal S$ is symmetric. Consequently,
$\mathcal S$ can transform a vector $\mathbf C$ sampled from a Maxwellian distribution to a vector $\mathbf c$ sampled from Eq.
\eqref{eq:esbgkdist}.
Here, an approach is used with an approximation of the transformation matrix $\mathcal S$ as described 
in previous studies \cite{gallis2011investigation, burt2006evaluation, Pfeiffer2018a} 
\begin{equation}
\mathcal S_{ij} = \delta_{ij}-\frac{1-\alpha Pr}{2\alpha Pr}\left[\frac{m}{k_{\mathrm{B}} T}\mathcal P_{ij}-\delta_{ij}\right].
\label{eq:Sconvert}
\end{equation}
In the context of particle methods, the required moments are evaluated with 
\begin{equation}
\mathcal{E}_s = \frac{1}{N_s-1} \sum_{p=1}^{N_s} \frac{m_s}{2} (\mathbf v_{p,s}-\mathbf u_s)^2,
\end{equation}
\begin{eqnarray}
     \mathcal P    && =\frac{1}{(\sum_{s=1}^M \sum_{p=1}^{N_s} m_s)(N_{\mathrm{total}}-1) }\nonumber\\
     && \times\sum_{s=1}^M \sum_{p=1}^{N_s} m_s (\mathbf v_{p,s}-\mathbf u) (\mathbf v_{p,s}-\mathbf u)^T,
\end{eqnarray}
with the particle number $N_s$ per species $s$ and the total particle number $N_{\mathrm{total}}=\sum_{s=1}^M N_s$. Here, the factors $\frac{1}{N_s-1}$ and $\frac{1}{N_{\mathrm{total}}-1}$ lead to an unbiasedness of the variance.
It is obvious that at least two particles per species are needed to calculate the species temperature. However, this leads to a bad estimation of the temperature and it is recommended to use at least 6-7 particles per species. If the case occurs that only one particle of a species is present in the cell, this species is skipped in the calculation of $\mathcal{E}_s$ and $T$.

\subsection{Energy and Momentum Conservation}
\label{sec:encon}
A detailed discussion of the possible energy and momentum conservation schemes for the particle BGK method can be found in \citet{Pfeiffer2018a}.
For the mixture model, the flow velocity and the thermal energy 
are determined before the collision ($\mathbf u$ and $E^{(\mathrm{th})}$) and for the provisional post-collision conditions ($\mathbf u^\dagger$ and $E^{\dagger,(\mathrm{th})}$):
\begin{align}
\mathbf u &= \frac{ \sum_{s=1}^M \sum_{p=1}^{N_s} m_s \mathbf v_{p,s}}{\sum_{s=1}^M \sum_{p=1}^{N_s}  m_s } \\
\mathbf E^{(\mathrm{th})} &=\sum_{s=1}^M n_s\mathcal{E}_s =\sum_{s=1}^M \sum_{p=1}^{N_s}  \frac{m_s (\mathbf v_{p,s} - \mathbf u)^2}{2}.
\end{align}
The final post-collision velocity $\mathbf v^*$ of every molecule (whether having undergone a relaxation or not) is then determined from the 
provisional post-collision velocity $\mathbf v^\dagger$ according to
\begin{equation}
\mathbf v^*=\mathbf u +(\mathbf v^\dagger- \mathbf u^\dagger)\sqrt{\frac{E^{(\mathrm{th})}}{E^{\dagger,(\mathrm{th})}}}.
\end{equation}

\section{Simulation Results}

\subsection{Reservoir Simulations}

As a first verification step, simple reservoir (or heat bath) simulations of argon-neon and argon-helium mixtures were performed. These simulations allow to verify the transient relaxation behavior as well as the equilibrium temperature. The gas mixture is initialized at different species temperatures ($T_{\mathrm{Ar,0}}=\SI{9000}{\kelvin}$, $T_{\mathrm{He/Ne,0}}=\SI{1000}{\kelvin}$, $n_{\mathrm{Ar/Ne/He}}=\SI{1e23}{\per\cubic\meter}$) in a single cell with perfect specular reflection at the boundary. To be able to compare the transient behavior, the time has to be normalized with the respective characteristic relaxation time $\tau_{\mathrm{c}}$. It corresponds to the point where the relaxation has progressed to $1/e$, for argon the characteristic temperature is $T_{\mathrm{c}}=\SI{6471.5}{\kelvin}$ and the determined relaxation times for the different methods and gases are summarized in \cref{tab:reservoir_relaxation_times}.

\begin{table}
\caption{Characteristic relaxation times $[\si{\second}]$ of argon from $T_{\mathrm{Ar,0}}=\SI{9000}{\kelvin}$ to $T_{\mathrm{c}}=\SI{6471.5}{\kelvin}$.}\label{tab:reservoir_relaxation_times}
  \begin{ruledtabular}\renewcommand*{\arraystretch}{1.4}
    \begin{tabular}{cccc}
      \multicolumn{2}{c}{Ar-Ne} & \multicolumn{2}{c}{Ar-He} \\
      DSMC &   ESBGK    &   DSMC    &   ESBGK    \\
      \hline
      \num{2.18689E-08} & \num{2.26103E-08} & \num{3.88548E-08} & \num{1.23369E-08}\\
    \end{tabular}
  \end{ruledtabular}
\end{table}

\begin{figure}\centering
  \subfloat[Argon-Neon]{\includegraphics{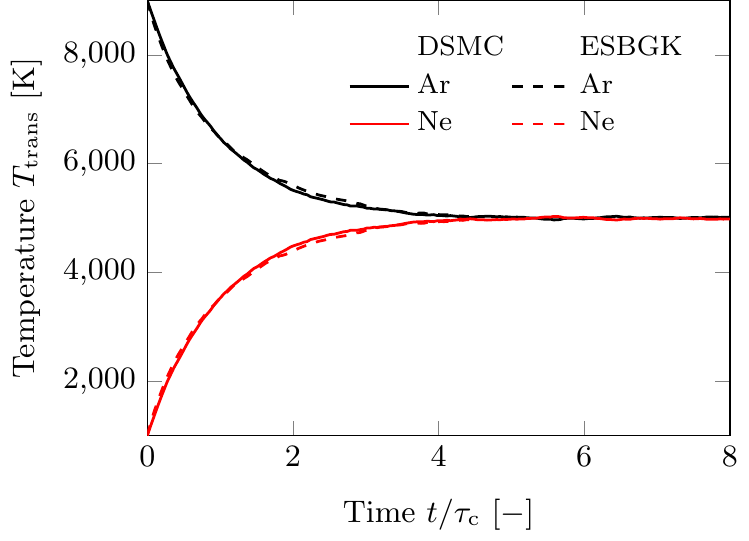}}
  
  \subfloat[Argon-Helium]{\includegraphics{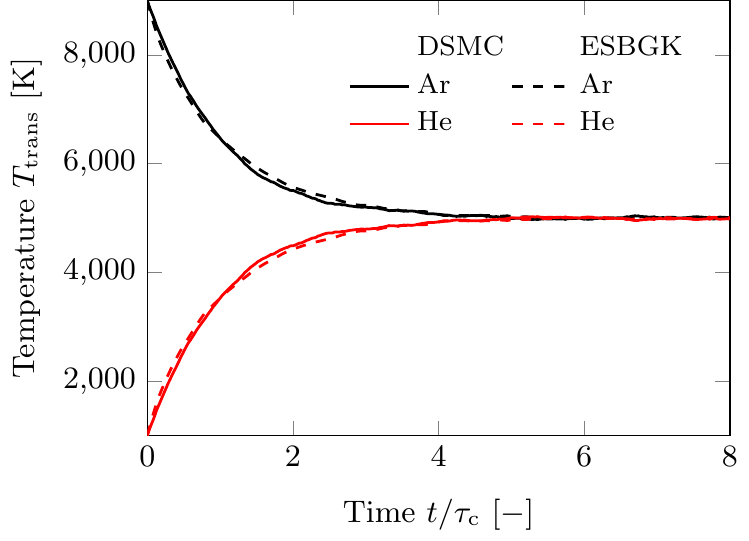}}
  \caption{Reservoir (50\%:50\% mixture): Comparison of species temperature towards thermal equilibrium between DSMC and ESBGK (Brull's model and Wilke's mixing rules). Time is normalized with the respective relaxation time $\tau_{\mathrm{c}}$.}\label{fig:reservoir}
\end{figure}

The simulation results using Wilke's mixing rules are shown in \cref{fig:reservoir} and agree well with DSMC in terms of the transient behavior as well as the final equilibrium temperature. Slightly better agreement can be seen for the argon-neon mixture due to the lower mass difference of $m_{\mathrm{Ar}}/m_{\mathrm{Ne}}\approx2$ as compared to $m_{\mathrm{Ar}}/m_{\mathrm{He}}\approx10$.

\subsection{Supersonic Couette Flow}
\label{sec:supercouette}
The second test case is a supersonic Couette flow. Supersonic conditions were chosen as they push the limits of the methods. Here, the different approaches, Wilke's mixing rules (denoted by Wilke) and the collision integrals (denoted by CollInt) for the determination of the transport coefficients are compared. The setup is one-dimensional with a height of \SI{1}{\meter} and 100 cells in the $y$-direction and a single cell in $x$ and $z$. The top and bottom boundaries have a velocity of $v_{\mathrm{top}}=\SI{350}{\metre\per\second}$ and $v_{\mathrm{bot}}=\SI{-350}{\metre\per\second}$, respectively. Additionally, diffuse reflection and complete thermal accommodation at a constant wall temperature of $T_{\mathrm{wall}}=\SI{273}{\kelvin}$ is assumed at the boundary. The boundaries in $x$- and $z$-direction are periodic, meaning that particles leaving on one side reappear on the other. The gas mixture is initialized at $v_0=0$, $T_0=\SI{273}{\kelvin}$, and $n = \SI{1.3e20}{\per\cubic\meter}$. After a transient phase, the stationary solution is utilized for comparison with the reference DSMC simulation.

\begin{figure}\centering
  \subfloat[Temperature]{\includegraphics{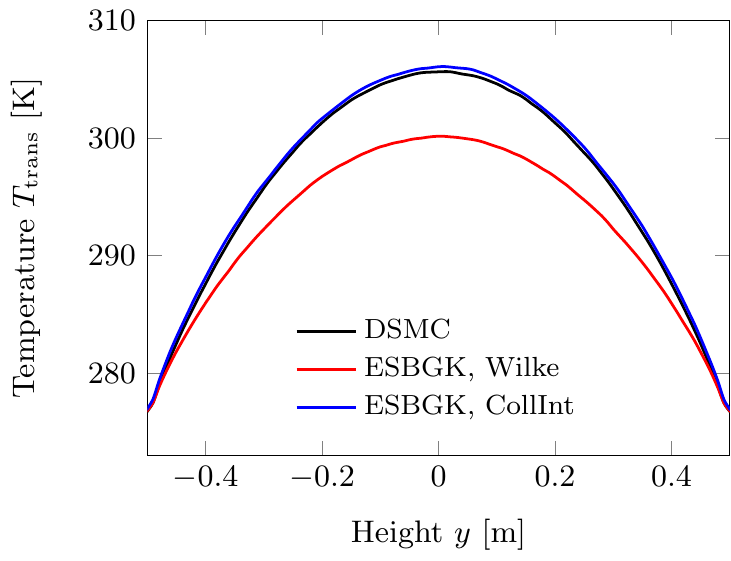}}

  \subfloat[Number density]{\includegraphics{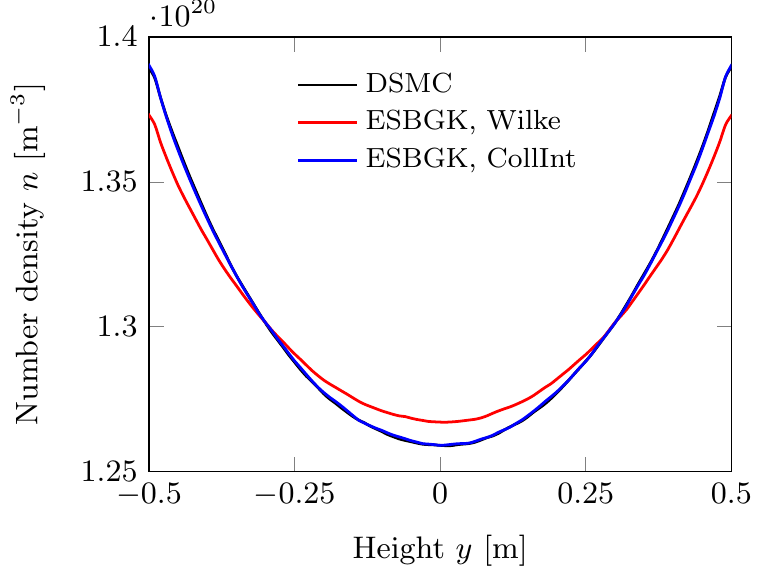}}
  \caption{Comparison of the stationary solution for a supersonic Couette flow for a 50\%-50\% argon-helium mixture.}\label{fig:couette_Ar_50_He_50}
\end{figure}

\begin{figure}\centering
  \subfloat[Temperature]{\includegraphics{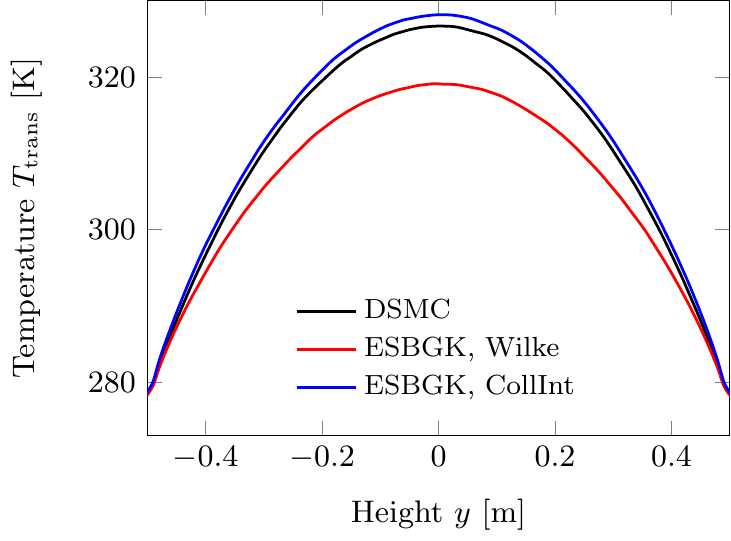}}

  \subfloat[Number density]{\includegraphics{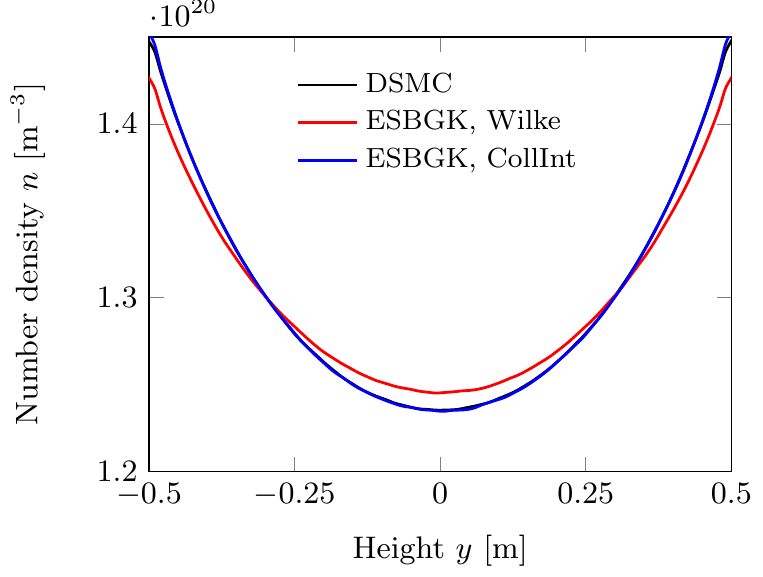}}
  \caption{Comparison of the stationary solution for a supersonic Couette flow for a 75\%-25\% argon-helium mixture.}\label{fig:couette_Ar_75_He_25}
\end{figure}

The simulation results of an argon-helium mixture, which represent a challenging case due to the relatively large mass difference, are shown in \cref{fig:couette_Ar_50_He_50} and \cref{fig:couette_Ar_75_He_25} for a $50\%$-$50\%$ and $75\%$-$25\%$ ratio, respectively. The results using Wilke's mixing rules show good agreement for the number density as well as translational temperature. For the latter, the deviation from the DSMC result is below $2\%$ for the $50\%$-$50\%$ mixture case and below $2.5\%$ for the $75\%$-$25\%$ case. Excellent agreement can be observed using the collision integrals for the number density and temperature, where the temperature deviation is below $0.2\%$ for the $50\%$-$50\%$ and below $0.4\%$ for the $75\%$-$25\%$ case.

\begin{figure}\centering
  \subfloat[Temperature]{\includegraphics{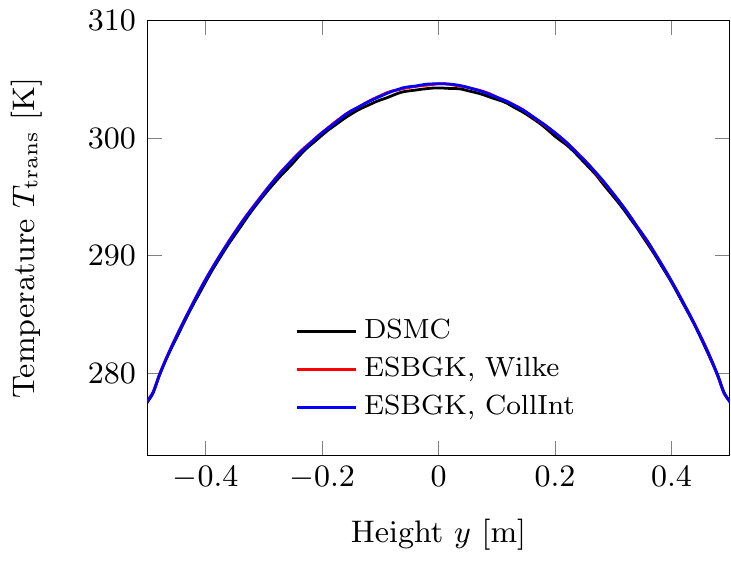}}
  
  \subfloat[Number density]{\includegraphics{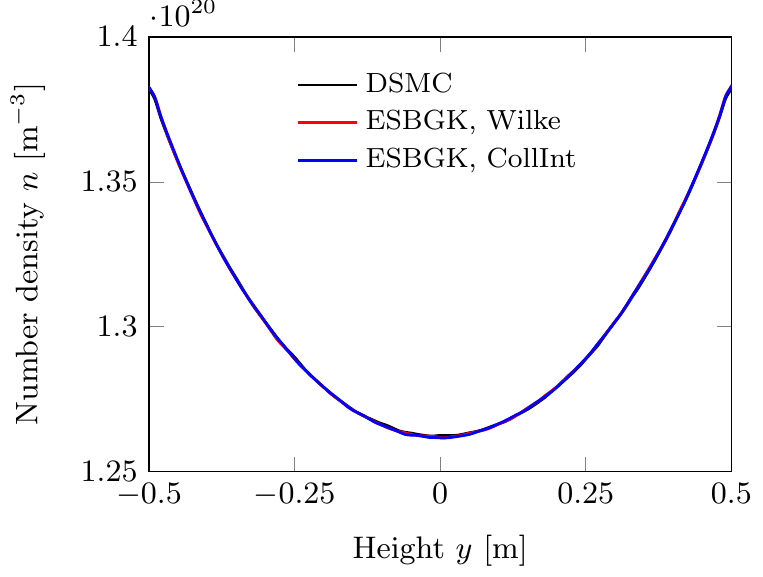}}
  \caption{Comparison of the stationary solution for a supersonic Couette flow for a 50\%-50\% nitrogen-oxygen mixture.}\label{fig:couette_N_50_O_50}
\end{figure}

The results of a nitrogen-oxygen mixture are shown in \cref{fig:couette_N_50_O_50}. They demonstrate that for lower mass ratios, Wilke's mixing rules achieve very good agreement with the DSMC result.

\subsection{70 Degree Blunted Cone}

\begin{figure}\centering
  \includegraphics[width=0.75\linewidth]{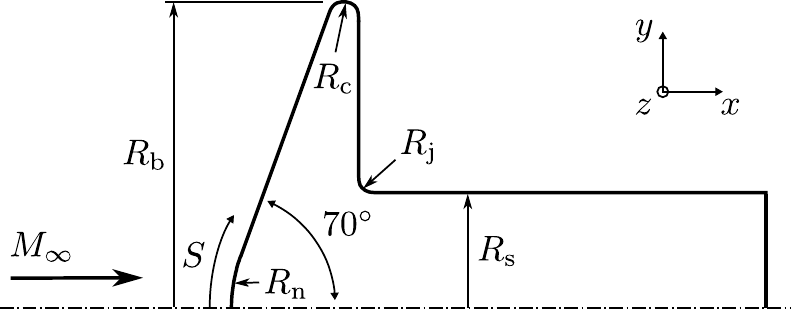}
  \caption{Geometry of the $70^\circ$ blunted cone. $R_{\mathrm{b}}=\SI{25.0}{\milli\meter}$, $R_{\mathrm{c}}=\SI{1.25}{\milli\meter}$, $R_{\mathrm{j}}=\SI{2.08}{\milli\meter}$, $R_{\mathrm{n}}=\SI{12.5}{\milli\meter}$, $R_{\mathrm{s}}=\SI{6.25}{\milli\meter}$. $S$ denotes the arc length along the surface.}\label{fig:70cone_geometry}
\end{figure}

The last verification case is the hypersonic flow around a 70$^\circ$ blunted cone. The geometry of the model, which is based on a wind-tunnel experiment, is shown in \cref{fig:70cone_geometry}. Axisymmetric simulations are performed, where the particle weighting factor increases with an increasing $y$. While the DSMC simulations require a particle "cloning"/deletion to ensure that the particle weights of colliding particles are similar, the ESBGK approach can handle different particle weights without additional particle manipulation. The surface of the blunted cone is diffusively reflective and with complete thermal accommodation at a constant wall temperature of $T_{\mathrm{w}}=\SI{300}{\kelvin}$. Three different test are performed with the inflow conditions shown in \cref{tab:cases} to investigate different compositions as well as mass ratios of the included atoms.

\begin{table}[htb!]
  \caption{Inflow conditions for $70^{\circ}$ blunted cone cases.\label{tab:cases}}
  \begin{tabular}{@{}l l l l l@{}}
      & $n_{\infty}$ [1/m$^3$] & $T_{\infty}$ [K] & $u_{\infty}$ [m/s] & Composition\\
    \hline
     Case 1 & $7.43\cdot 10^{20}$ & 13.3 & 1502.57 & $\nicefrac{3}{4}$N-$\nicefrac{1}{4}$O \\
     Case 2 & $7.43\cdot 10^{20}$ & 13.3 & 1502.57 & $\nicefrac{3}{4}$Ar-$\nicefrac{1}{4}$He \\
     Case 3 & $2.4\cdot 10^{21}$ & 13.3 & 1502.57 & $\nicefrac{1}{3}$N-$\nicefrac{1}{3}$O-$\nicefrac{1}{3}$Ar \\
  \end{tabular}
\end{table}

\subsubsection{Case 1} 
\begin{figure}
\centering
\includegraphics{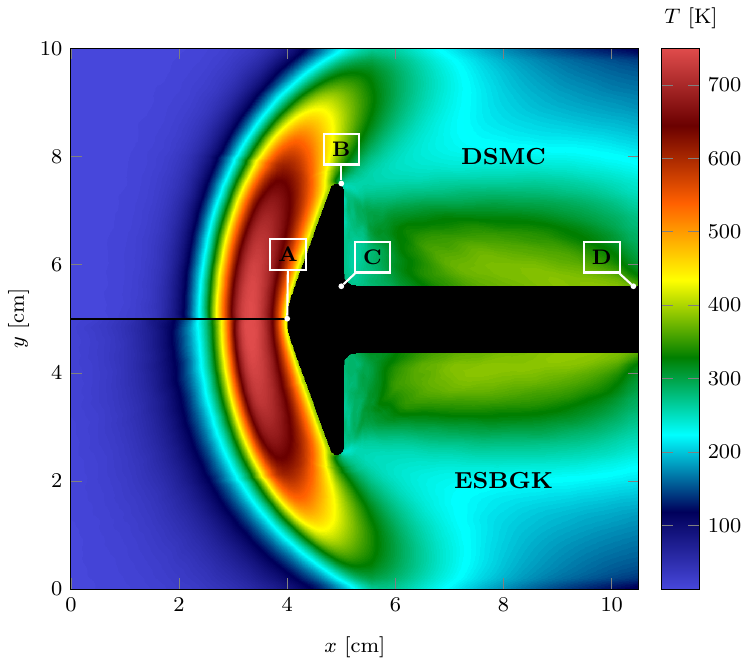}
\caption{$70^\circ$ blunted cone, Case 1: Temperature plots of the flow field using DSMC and ESBGK.}
\label{Fig:compTtrans}
\end{figure}

The first test is performed with an atomic nitrogen-oxygen mixture at a $75\%$-$25\%$ ratio. A comparison of the mean translational temperature of the gas mixture using DSMC and ESBGK
is depicted in \cref{Fig:compTtrans}. The results using Wilke's mixing rules or collision integrals are almost identical in this case and are therefore not shown in this comparison. The ESBGK model predicts an early onset of the temperature increase compared with DSMC, which results in slightly wider shock profiles. However, the overall agreement for with the DSMC result is very good.

\begin{figure}
\centering
\includegraphics{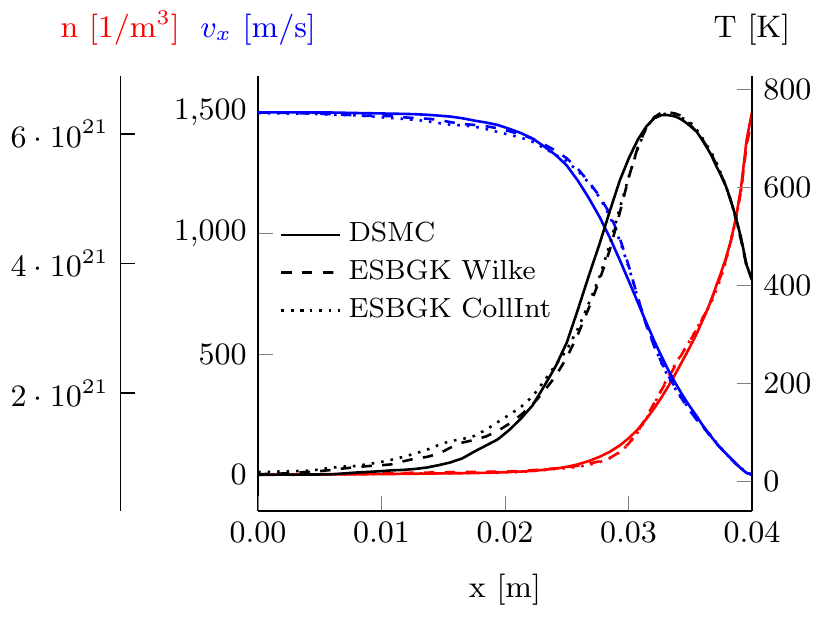}
\caption{$70^\circ$ blunted cone, Case 1: Mixture mean values of gas temperature, velocity in x-direction and number density along stagnation stream line using DSMC and ESBGK.}
\label{Fig:set1mean}
\end{figure}

\begin{figure}
\centering
\includegraphics{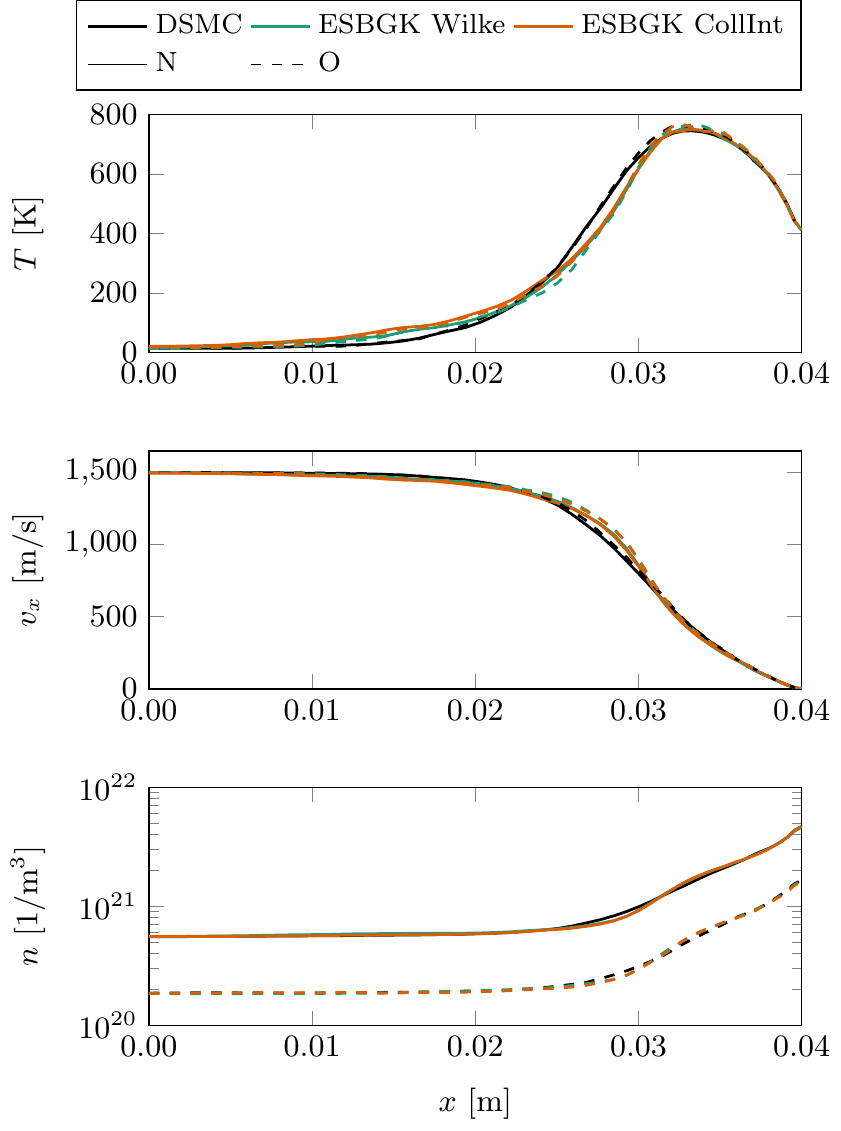}
\caption{$70^\circ$ blunted cone, Case 1: Species temperatures, velocities in x-direction, and number densities along the stagnation stream line using DSMC and ESBGK.}
\label{Fig:set1comp}
\end{figure}

The simulation results of the mean flow variables of the mixture as well as the species flow variables over the stagnation stream line are shown in \cref{Fig:set1mean} and \cref{Fig:set1comp}, respectively. The overall agreement for both methods with the DSMC result is very good. Again, small differences in the temperature can be observed during the onset of the shock, however, the agreement in the post-shock region is excellent. 

\begin{figure}
\centering
\includegraphics{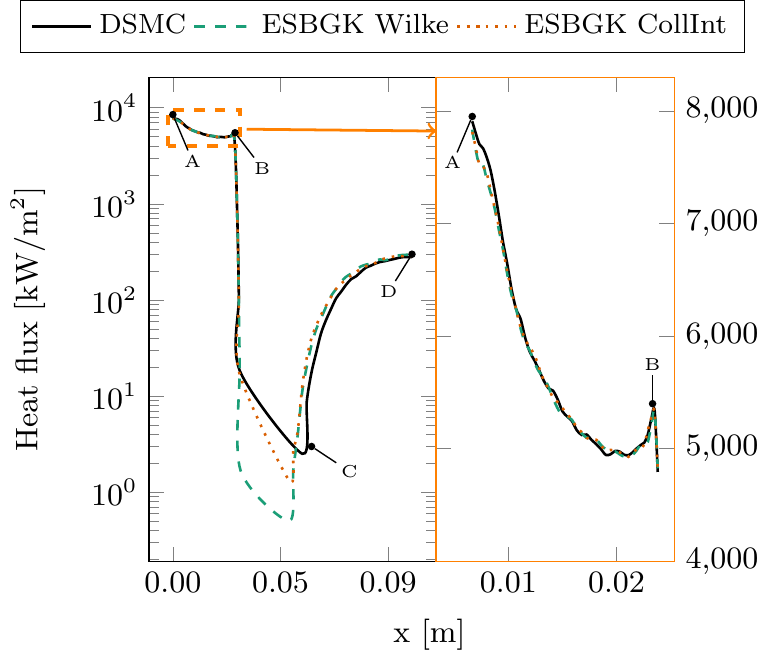}
\caption{$70^\circ$ blunted cone, Case 1: Heat flux on the surface.}
\label{Fig:set1heat}
\end{figure}

\begin{figure}
\centering
\includegraphics{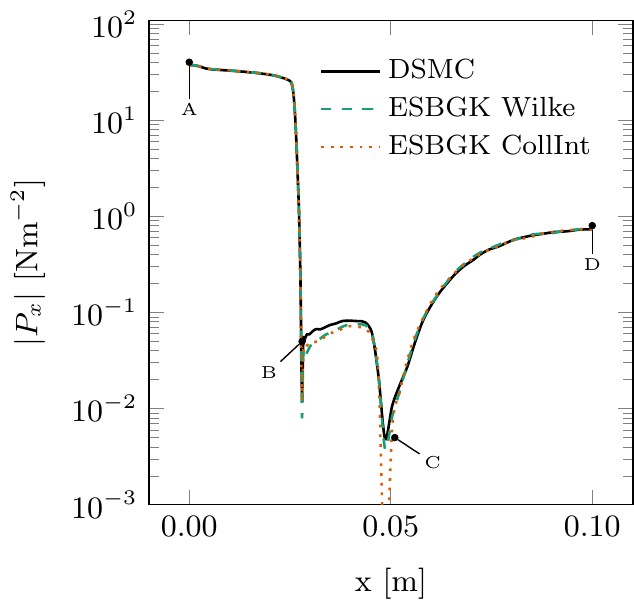}
\caption{$70^\circ$ blunted cone, Case 1: Pressure x direction on the surface.\label{fig:forcecase1}}
\end{figure}

The heat flux and pressure on the surface of the cone are depicted in \cref{Fig:set1heat} and \cref{fig:forcecase1} with the points \{A,B,C,D\} corresponding to the points depicted in \cref{Fig:compTtrans}. Both show excellent agreement on the flow-facing heat shield as well as the sting further downstream.

\subsubsection{Case 2} 
\begin{figure}
\centering
\includegraphics{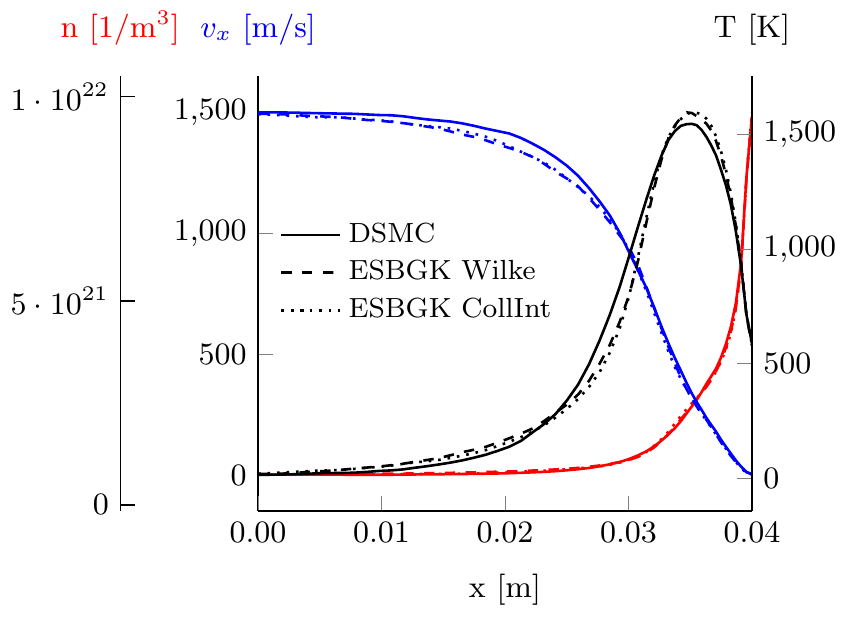}
\caption{$70^\circ$ blunted cone, Case 2: Mixture mean values of gas temperature, velocity in x-direction and number density along stagnation stream line.}
\label{Fig:set2mean}
\end{figure}

\begin{figure}
\centering
\includegraphics{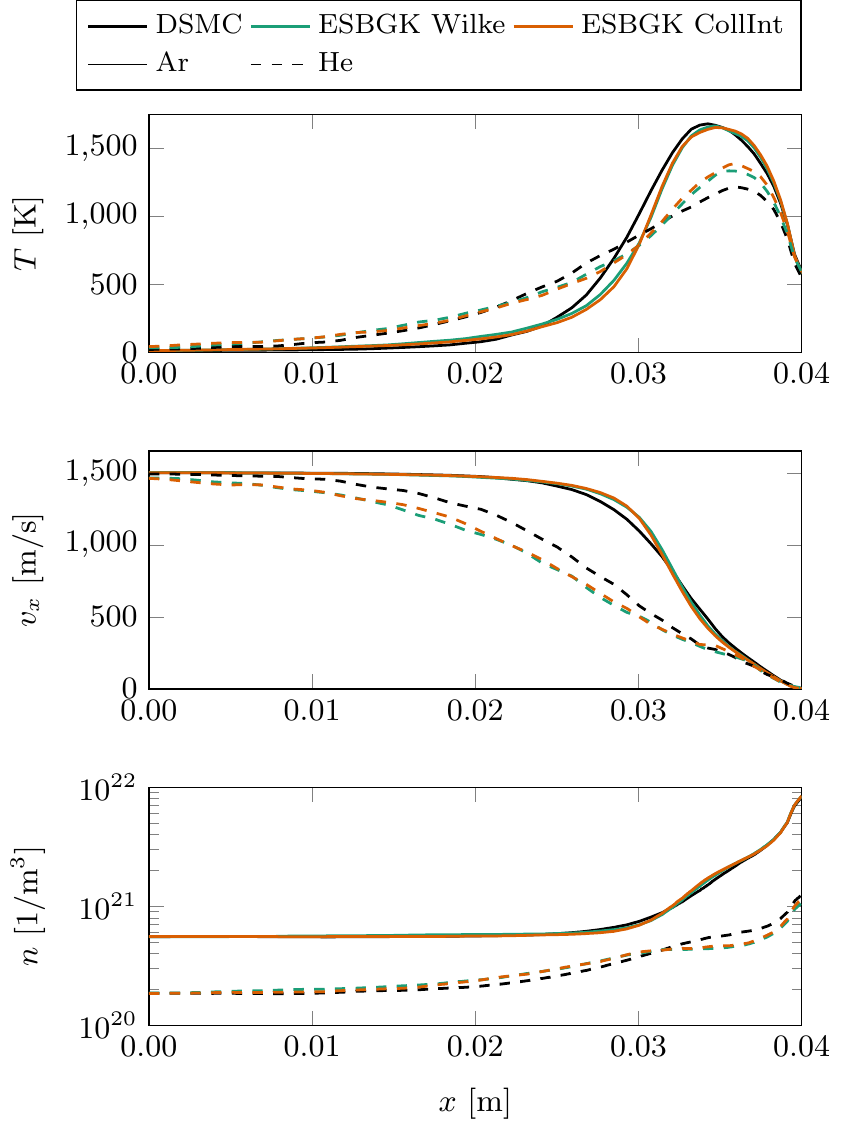}
\caption{$70^\circ$ blunted cone, Case 2: Species temperatures, velocities in x-direction, and number densities along the stagnation stream line using DSMC and ESBGK.}
\label{Fig:set2comp}
\end{figure}

Case 2 is performed with an argon-helium mixture at a $75\%$-$25\%$ ratio. This mixture has a significant higher mass ratio $\nicefrac{m_{\mathrm{Ar}}}{m_{\mathrm{He}}}\approx 10$ compared with the first case $\nicefrac{m_{\mathrm{O}}}{m_{\mathrm{N}}}\approx 1.14$. Therefore a greater difference between Wilke's mixing rules and the collision integral approach is expected here. However, the 
results of the mean flow variables of the mixture (\cref{Fig:set2mean}) as well as the species flow variables (\cref{Fig:set2comp}) over the stagnation stream line are almost identical for both models. Furthermore, the agreement of both ESBGK models with DSMC regarding the mean flow variables in \cref{Fig:set2mean} is very good, the deviations are only slightly larger than in Case 1. 

The differences between the ESBGK models and DSMC are more pronounced for the individual species in \cref{Fig:set2comp}. Nevertheless, the deviation is still relatively small for the large mass ratio. We expect that this difference can be further reduced with more sophisticated models, e.g. by \citet{Klingenberg2018b, Todorova2019}. These will be implemented and compared in the future.

\begin{figure}
\centering
\includegraphics{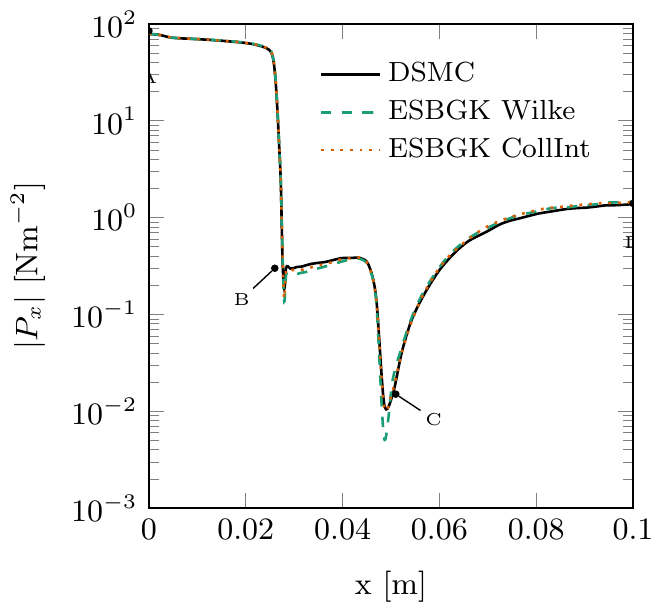}
\caption{$70^\circ$ blunted cone, Case 2: Pressure x direction on the surface.\label{fig:forcecase2}}
\end{figure}

\begin{figure}
\centering
\includegraphics{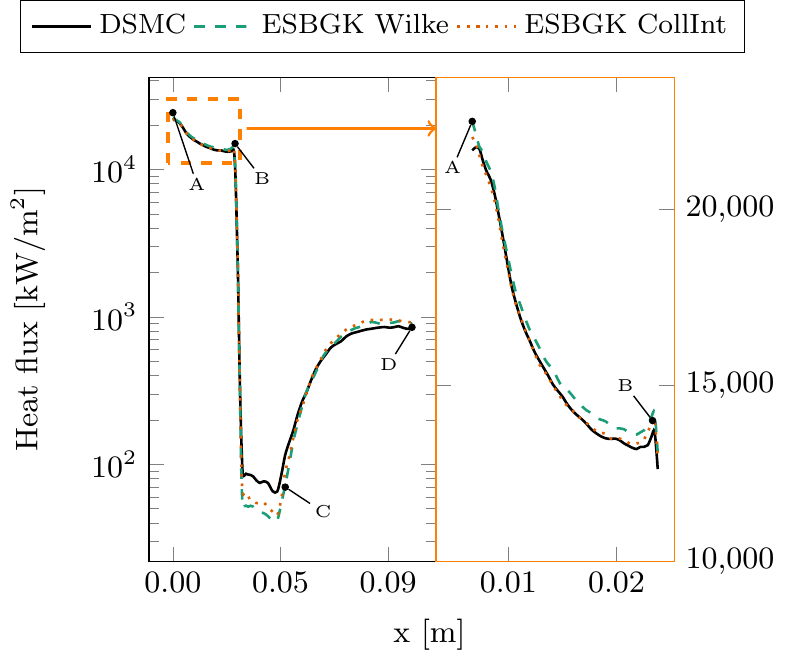}
\caption{$70^\circ$ blunted cone, Case 2: Heat flux on the surface.}
\label{Fig:set2heat}
\end{figure}

The pressure on the surface depicted in \cref{fig:forcecase2} is almost identical for DSMC and both ESBGK models. The heat flux, however, shows a difference between the two ESBGK models, where the collision integral ESBGK model matches the DSMC result very well on the flow-facing heat shield while Wilke's mixing rules show a slight deviation from the DSMC result as shown in \cref{Fig:set2heat}. This result was to be expected with regard to the results of the supersonic Couette flow of \cref{sec:supercouette}.

\subsubsection{Case 3} 

\begin{figure}
\centering
\includegraphics{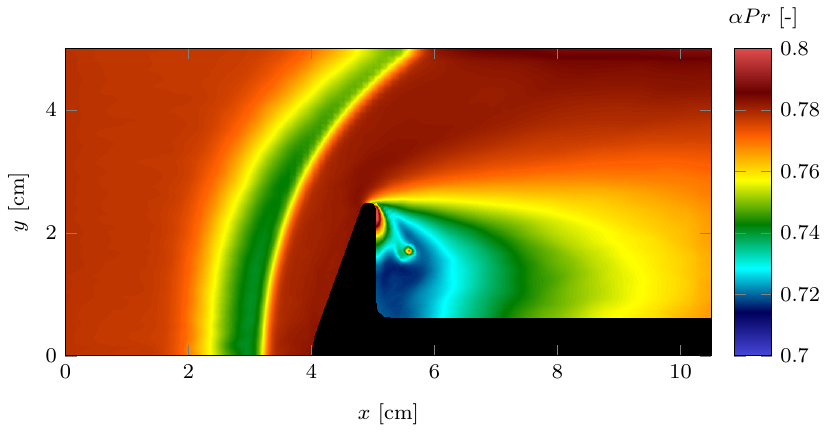}
\caption{$70^\circ$ blunted cone, Case 3: Plot of $\alpha Pr$ for the collision integral model.\label{fig:prandtl}}
\end{figure}

The third case is performed with an equal ratio nitrogen-oxygen-argon mixture with mass ratios of $\nicefrac{m_{\mathrm{Ar}}}{m_{\mathrm{N}}}\approx\nicefrac{m_{\mathrm{Ar}}}{m_{\mathrm{O}}}\approx 2.6$. For this case, the product $\alpha Pr$ is depicted in \cref{fig:prandtl} to illustrate the changing Prandtl number of the mixture in the flow. The factor $\alpha Pr$ is ranging between 0.7 and 0.8 in this example.

\begin{figure}
\centering
\includegraphics{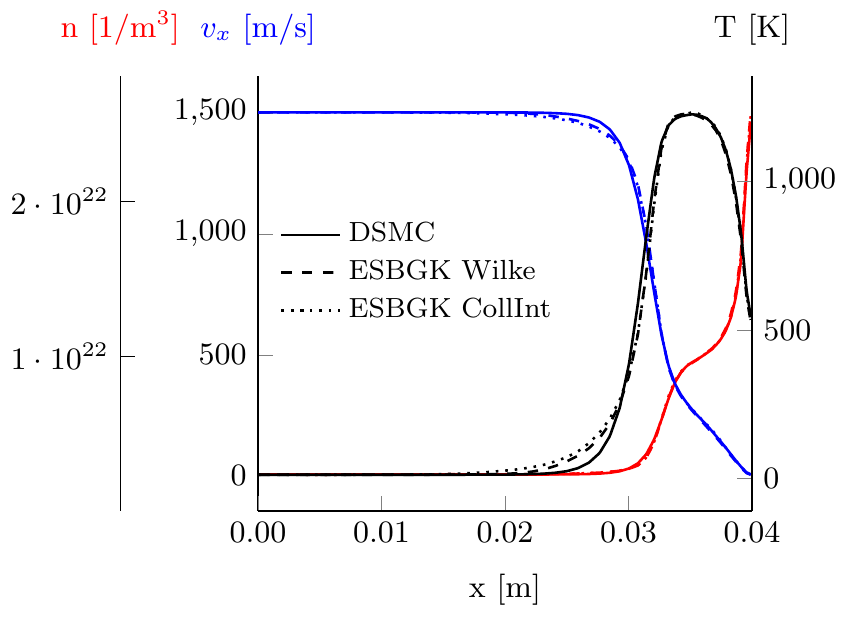}
\caption{$70^\circ$ blunted cone, Case 3: Mixture mean values of gas temperature, velocity in x-direction and number density along stagnation stream line.}
\label{Fig:set3mean}
\end{figure}

\begin{figure}
\centering
\includegraphics{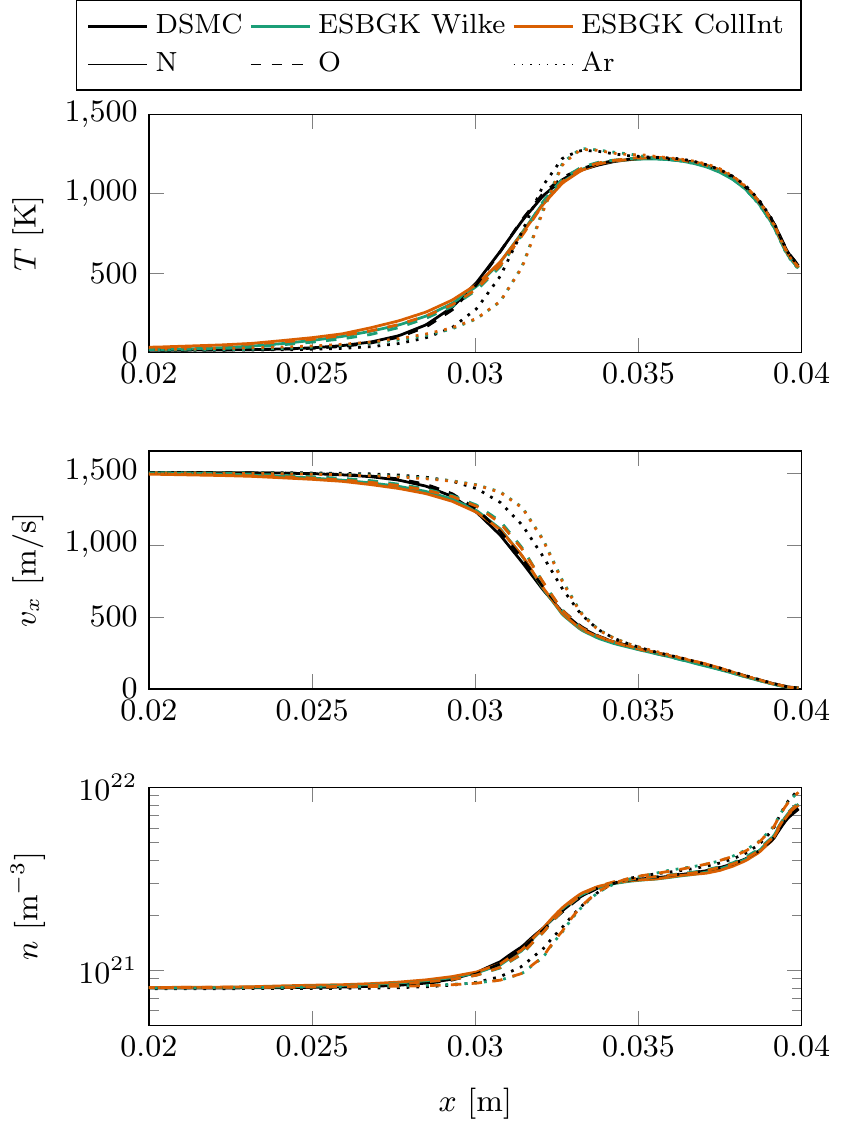}
\caption{$70^\circ$ blunted cone, Case 3: Species temperatures, velocities in x-direction, and number densities along the stagnation stream line using DSMC and ESBGK.}
\label{Fig:set3comp}
\end{figure}

Again, the results of the mean flow variables of the mixture (\cref{Fig:set3mean}) as well as the species flow variables (\cref{Fig:set3comp}) over the stagnation stream line show a very good agreement between DSMC and both ESBGK models. The DSMC temperature curve of argon as the heaviest species is well reproduced with the ESBGK modeling as depicted in \cref{Fig:set3comp}.

\begin{figure}
\centering
\includegraphics{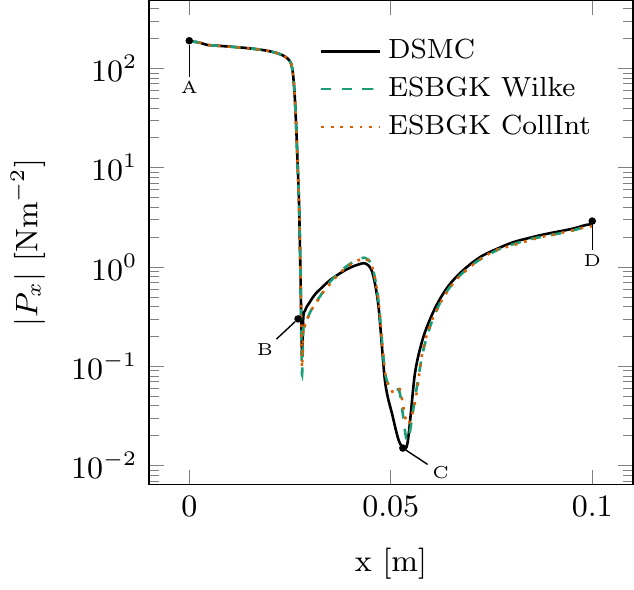}
\caption{$70^\circ$ blunted cone, Case 3: Pressure in $x$-direction on the surface.\label{fig:forcecase3}}
\end{figure}

\begin{figure}
\centering
\includegraphics{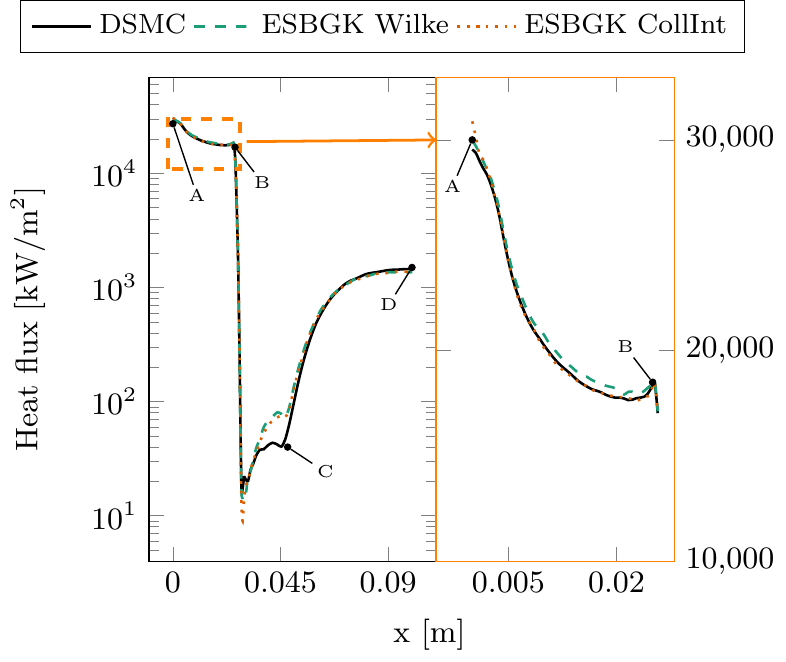}
\caption{$70^\circ$ blunted cone, Case 3: Heat flux on the surface.}
\label{Fig:set3heat}
\end{figure}

The pressure on the surface depicted in \cref{fig:forcecase3} is practically identical again for DSMC and ESBGK. The heat flux of the collision integral ESBGK model matches the DSMC result better than Wilke's mixing rules on the flow-facing heat shield as depicted in \cref{Fig:set3heat}. But again, the result with Wilke's mixing rules is also quite good.

A performance comparison for 100 iterations between DSMC as well as ESBGK with Wilke's mixing rules and the collision integral approach is given in \cref{tab:cputime} for the third case.
In cases 1 and 2 the computing time between DSMC and ESBGK was almost the same due to the degree of rarefication. This has already been described in detail by \citet{Pfeiffer2018a,Pfeiffer2019a} and is expected that DSMC will be faster than ESBGK, when the number of collisions is low. In the third test case, however, the density was slightly increased and as shown in \cref{tab:cputime}, the ESBGK method outperforms DSMC. This is partly because DSMC needs more particles to resolve the mean free path, and partly because DSMC requires a smaller time step to resolve the collision frequency in this case. As a result, DSMC takes about 14 times longer than ESBGK to achieve the same simulation time for the third test case. The difference in simulation time between ESBGK with Wilke's model and the collision integral model is acceptable considering the increased accuracy in the simulation results.

\begin{table}
  \begin{center}
\footnotesize
\def~{\hphantom{0}}
  \begin{tabular}{l C{1.5cm} C{2cm} C{2cm}}
   &  Time step $\Delta t$ [s] &  CPU Time / 300 iterations [s] & CPU Time / $4.5\cdot 10^{-5}\,\mathrm{s}$ Simulation time [s] \\
  \hline
  DSMC & $2\cdot 10^{-8}$  & 112 & 840 \\
  ESBGK Wilke & $1.5\cdot 10^{-7}$ & 55 & 55 \\
  ESBGK CollInt & $1.5\cdot 10^{-7}$ & 60 & 60
  \end{tabular}
  \caption{\label{tab:cputime} Comparison of CPU time between DSMC ($N^{\mathrm{DSMC}}_{\mathrm{part}}=\num{6.1e6}$) as well as ESBGK ($N^{\mathrm{BGK}}_{\mathrm{part}}=N^{\mathrm{DSMC}}_{\mathrm{part}}/2$) with Wilke's mixing rules and the collision integral model for Case 3. The CPU time is the time per node with 40 cores on an Intel Xeon Platinum 8160 CPU @ 2.10GHz.}
  \end{center}
\end{table}

\section{Conclusion}

Multi-species modeling for atomic species in the particle-based ellipsoidal statistical Bhatnagar-Gross-Krook model is implemented using Brull's model \cite{Brull2014}. For the determination of the transport coefficients two approaches have been implemented. The first relies on Wilke's mixture rules to determine the mixture properties and the second calculates them from collision integrals (for the Variable Hard Sphere model). The implementation is verified with reservoir test cases, a supersonic Couette flow test case and the hypersonic flow around a $70^{\circ}$ blunted cone at different free-stream conditions, and the results are compared with the DSMC method. The collision integral model offers the best agreement overall, and especially when the mass ratio is high (e.g. in an argon-helium mixture) compared to Wilke's mixing rules. However, Wilke's mixing rules show good agreement for smaller mass ratios as demonstrated in supersonic Couette and the $70^{\circ}$ blunted cone test cases. Although the calculation of the transport coefficients through the collision integrals is more complex, a first performance comparison of the computational effort suggests that the increase is below $10\%$ compared to Wilke's mixing rules.

The next steps in the development include the investigation of more advanced models for the extension of the kinetic equations (e.g. \citet{Todorova2019} and \citet{Klingenberg2018b}) and the extension of the current models to diatomic molecules. Looking further ahead, modeling of chemical reactions shall allow the bidirectional coupling with the DSMC method for a multitude of applications such as the simulation of atmospheric entry maneuvers and in-space propulsion.

\section*{Acknowledgments}

The authors gratefully acknowledge the Deutsche Forschungsgemeinschaft (DFG) for funding this research within the project “Partikelverfahren mit Strahlungslöser zur Simulation hochenthalper Nichtgleichgewichts-Plasmen” (project number 93159129). Part of the work was conducted under a program of and funded by the European Space Agency.
 
\bibliography{mybibfile}

\begin{thebibliography}{31}%
\makeatletter
\providecommand \@ifxundefined [1]{%
 \@ifx{#1\undefined}
}%
\providecommand \@ifnum [1]{%
 \ifnum #1\expandafter \@firstoftwo
 \else \expandafter \@secondoftwo
 \fi
}%
\providecommand \@ifx [1]{%
 \ifx #1\expandafter \@firstoftwo
 \else \expandafter \@secondoftwo
 \fi
}%
\providecommand \natexlab [1]{#1}%
\providecommand \enquote  [1]{``#1''}%
\providecommand \bibnamefont  [1]{#1}%
\providecommand \bibfnamefont [1]{#1}%
\providecommand \citenamefont [1]{#1}%
\providecommand \href@noop [0]{\@secondoftwo}%
\providecommand \href [0]{\begingroup \@sanitize@url \@href}%
\providecommand \@href[1]{\@@startlink{#1}\@@href}%
\providecommand \@@href[1]{\endgroup#1\@@endlink}%
\providecommand \@sanitize@url [0]{\catcode `\\12\catcode `\$12\catcode
  `\&12\catcode `\#12\catcode `\^12\catcode `\_12\catcode `\%12\relax}%
\providecommand \@@startlink[1]{}%
\providecommand \@@endlink[0]{}%
\providecommand \url  [0]{\begingroup\@sanitize@url \@url }%
\providecommand \@url [1]{\endgroup\@href {#1}{\urlprefix }}%
\providecommand \urlprefix  [0]{URL }%
\providecommand \Eprint [0]{\href }%
\providecommand \doibase [0]{http://dx.doi.org/}%
\providecommand \selectlanguage [0]{\@gobble}%
\providecommand \bibinfo  [0]{\@secondoftwo}%
\providecommand \bibfield  [0]{\@secondoftwo}%
\providecommand \translation [1]{[#1]}%
\providecommand \BibitemOpen [0]{}%
\providecommand \bibitemStop [0]{}%
\providecommand \bibitemNoStop [0]{.\EOS\space}%
\providecommand \EOS [0]{\spacefactor3000\relax}%
\providecommand \BibitemShut  [1]{\csname bibitem#1\endcsname}%
\let\auto@bib@innerbib\@empty
\bibitem [{\citenamefont {Bird}(1994)}]{Bird1994}%
  \BibitemOpen
  \bibfield  {author} {\bibinfo {author} {\bibfnamefont {G.~A.}\ \bibnamefont
  {Bird}},\ }\href@noop {} {\emph {\bibinfo {title} {{Molecular Gas Dynamics
  and the Direct Simulation of Gas Flows}}}},\ \bibinfo {edition} {2nd}\ ed.\
  (\bibinfo  {publisher} {Oxford University Press},\ \bibinfo {address} {New
  York},\ \bibinfo {year} {1994})\BibitemShut {NoStop}%
\bibitem [{\citenamefont {Zhang}\ \emph {et~al.}(2019)\citenamefont {Zhang},
  \citenamefont {John}, \citenamefont {Pfeiffer}, \citenamefont {Fei},\ and\
  \citenamefont {Wen}}]{zhang2019particle}%
  \BibitemOpen
  \bibfield  {author} {\bibinfo {author} {\bibfnamefont {J.}~\bibnamefont
  {Zhang}}, \bibinfo {author} {\bibfnamefont {B.}~\bibnamefont {John}},
  \bibinfo {author} {\bibfnamefont {M.}~\bibnamefont {Pfeiffer}}, \bibinfo
  {author} {\bibfnamefont {F.}~\bibnamefont {Fei}}, \ and\ \bibinfo {author}
  {\bibfnamefont {D.}~\bibnamefont {Wen}},\ }\bibfield  {title} {\enquote
  {\bibinfo {title} {Particle-based hybrid and multiscale methods for
  nonequilibrium gas flows},}\ }\href@noop {} {\bibfield  {journal} {\bibinfo
  {journal} {Advances in Aerodynamics}\ }\textbf {\bibinfo {volume} {1}},\
  \bibinfo {pages} {12} (\bibinfo {year} {2019})}\BibitemShut {NoStop}%
\bibitem [{\citenamefont {Pfeiffer}, \citenamefont {Mirza},\ and\ \citenamefont
  {Nizenkov}(2019)}]{Pfeiffer2019a}%
  \BibitemOpen
  \bibfield  {author} {\bibinfo {author} {\bibfnamefont {M.}~\bibnamefont
  {Pfeiffer}}, \bibinfo {author} {\bibfnamefont {A.}~\bibnamefont {Mirza}}, \
  and\ \bibinfo {author} {\bibfnamefont {P.}~\bibnamefont {Nizenkov}},\
  }\bibfield  {title} {\enquote {\bibinfo {title} {{Evaluation of
  particle-based continuum methods for a coupling with the direct simulation
  Monte Carlo method based on a nozzle expansion}},}\ }\href {\doibase
  10.1063/1.5098085} {\bibfield  {journal} {\bibinfo  {journal} {Physics of
  Fluids}\ }\textbf {\bibinfo {volume} {31}},\ \bibinfo {pages} {073601}
  (\bibinfo {year} {2019})}\BibitemShut {NoStop}%
\bibitem [{\citenamefont {Bhatnagar}, \citenamefont {Gross},\ and\
  \citenamefont {Krook}(1954)}]{bhatnagar1954model}%
  \BibitemOpen
  \bibfield  {author} {\bibinfo {author} {\bibfnamefont {P.~L.}\ \bibnamefont
  {Bhatnagar}}, \bibinfo {author} {\bibfnamefont {E.~P.}\ \bibnamefont
  {Gross}}, \ and\ \bibinfo {author} {\bibfnamefont {M.}~\bibnamefont
  {Krook}},\ }\bibfield  {title} {\enquote {\bibinfo {title} {{A model for
  collision processes in gases. I. Small amplitude processes in charged and
  neutral one-component systems}},}\ }\href@noop {} {\bibfield  {journal}
  {\bibinfo  {journal} {Physical review}\ }\textbf {\bibinfo {volume} {94}},\
  \bibinfo {pages} {511} (\bibinfo {year} {1954})}\BibitemShut {NoStop}%
\bibitem [{\citenamefont {Holway}(1966)}]{Holway1966}%
  \BibitemOpen
  \bibfield  {author} {\bibinfo {author} {\bibfnamefont {L.~H.}\ \bibnamefont
  {Holway}},\ }\bibfield  {title} {\enquote {\bibinfo {title} {{New statistical
  models for kinetic theory: Methods of construction}},}\ }\href {\doibase
  10.1063/1.1761920} {\bibfield  {journal} {\bibinfo  {journal} {Physics of
  Fluids}\ }\textbf {\bibinfo {volume} {9}},\ \bibinfo {pages} {1658--1673}
  (\bibinfo {year} {1966})}\BibitemShut {NoStop}%
\bibitem [{\citenamefont {Shakhov}(1968)}]{Shakhov1968}%
  \BibitemOpen
  \bibfield  {author} {\bibinfo {author} {\bibfnamefont {E.~M.}\ \bibnamefont
  {Shakhov}},\ }\bibfield  {title} {\enquote {\bibinfo {title} {{Generalization
  of the Krook kinetic relaxation equation}},}\ }\href {\doibase
  10.1007/BF01029546} {\bibfield  {journal} {\bibinfo  {journal} {Fluid
  Dynamics}\ }\textbf {\bibinfo {volume} {3}},\ \bibinfo {pages} {95--96}
  (\bibinfo {year} {1968})}\BibitemShut {NoStop}%
\bibitem [{Note1()}]{Note1}%
  \BibitemOpen
  \bibinfo {note} {PICLas is a flexible particle-based plasma simulation suite.
  Available online at https://github.com/piclas-framework/piclas.}\BibitemShut
  {Stop}%
\bibitem [{\citenamefont {Pfeiffer}(2018{\natexlab{a}})}]{Pfeiffer2018a}%
  \BibitemOpen
  \bibfield  {author} {\bibinfo {author} {\bibfnamefont {M.}~\bibnamefont
  {Pfeiffer}},\ }\bibfield  {title} {\enquote {\bibinfo {title}
  {{Particle-based fluid dynamics: Comparison of different
  Bhatnagar-Gross-Krook models and the direct simulation Monte Carlo method for
  hypersonic flows}},}\ }\href {\doibase 10.1063/1.5042016} {\bibfield
  {journal} {\bibinfo  {journal} {Physics of Fluids}\ }\textbf {\bibinfo
  {volume} {30}},\ \bibinfo {pages} {106106} (\bibinfo {year}
  {2018}{\natexlab{a}})}\BibitemShut {NoStop}%
\bibitem [{\citenamefont {Pfeiffer}(2018{\natexlab{b}})}]{Pfeiffer2018b}%
  \BibitemOpen
  \bibfield  {author} {\bibinfo {author} {\bibfnamefont {M.}~\bibnamefont
  {Pfeiffer}},\ }\bibfield  {title} {\enquote {\bibinfo {title} {{Extending the
  particle ellipsoidal statistical Bhatnagar-Gross-Krook method to diatomic
  molecules including quantized vibrational energies}},}\ }\href {\doibase
  10.1063/1.5054961} {\bibfield  {journal} {\bibinfo  {journal} {Physics of
  Fluids}\ }\textbf {\bibinfo {volume} {30}},\ \bibinfo {pages} {116103}
  (\bibinfo {year} {2018}{\natexlab{b}})}\BibitemShut {NoStop}%
\bibitem [{\citenamefont {Pfeiffer}, \citenamefont {Nizenkov},\ and\
  \citenamefont {Fasoulas}(2019)}]{Pfeiffer2019c}%
  \BibitemOpen
  \bibfield  {author} {\bibinfo {author} {\bibfnamefont {M.}~\bibnamefont
  {Pfeiffer}}, \bibinfo {author} {\bibfnamefont {P.}~\bibnamefont {Nizenkov}},
  \ and\ \bibinfo {author} {\bibfnamefont {S.}~\bibnamefont {Fasoulas}},\
  }\bibfield  {title} {\enquote {\bibinfo {title} {{Extension of particle-based
  BGK models to polyatomic species in hypersonic flow around a flat-faced
  cylinder}},}\ }\href {\doibase 10.1063/1.5119596} {\bibfield  {journal}
  {\bibinfo  {journal} {AIP Conference Proceedings}\ }\textbf {\bibinfo
  {volume} {2132}},\ \bibinfo {pages} {100001} (\bibinfo {year}
  {2019})}\BibitemShut {NoStop}%
\bibitem [{\citenamefont {Andries}, \citenamefont {Aoki},\ and\ \citenamefont
  {Perthame}(2002)}]{Andries2002}%
  \BibitemOpen
  \bibfield  {author} {\bibinfo {author} {\bibfnamefont {P.}~\bibnamefont
  {Andries}}, \bibinfo {author} {\bibfnamefont {K.}~\bibnamefont {Aoki}}, \
  and\ \bibinfo {author} {\bibfnamefont {B.}~\bibnamefont {Perthame}},\
  }\bibfield  {title} {\enquote {\bibinfo {title} {{A Consistent BGK-Type Model
  for Gas Mixtures}},}\ }\href@noop {} {\bibfield  {journal} {\bibinfo
  {journal} {Journal of Statistical Physics}\ }\textbf {\bibinfo {volume}
  {106}},\ \bibinfo {pages} {993--1018} (\bibinfo {year} {2002})}\BibitemShut
  {NoStop}%
\bibitem [{\citenamefont {Klingenberg}\ and\ \citenamefont
  {Pirner}(2018)}]{Klingenberg2018a}%
  \BibitemOpen
  \bibfield  {author} {\bibinfo {author} {\bibfnamefont {C.}~\bibnamefont
  {Klingenberg}}\ and\ \bibinfo {author} {\bibfnamefont {M.}~\bibnamefont
  {Pirner}},\ }\bibfield  {title} {\enquote {\bibinfo {title} {{Existence ,
  uniqueness and positivity of solutions for BGK models for mixtures}},}\
  }\href {\doibase 10.1016/j.jde.2017.09.019} {\bibfield  {journal} {\bibinfo
  {journal} {Journal of Differential Equations}\ }\textbf {\bibinfo {volume}
  {264}},\ \bibinfo {pages} {702--727} (\bibinfo {year} {2018})}\BibitemShut
  {NoStop}%
\bibitem [{\citenamefont {Klingenberg}, \citenamefont {Pirner},\ and\
  \citenamefont {Puppo}(2018)}]{Klingenberg2018b}%
  \BibitemOpen
  \bibfield  {author} {\bibinfo {author} {\bibfnamefont {C.}~\bibnamefont
  {Klingenberg}}, \bibinfo {author} {\bibfnamefont {M.}~\bibnamefont {Pirner}},
  \ and\ \bibinfo {author} {\bibfnamefont {G.}~\bibnamefont {Puppo}},\
  }\bibfield  {title} {\enquote {\bibinfo {title} {{Kinetic ES-BGK Models for a
  Multi-component Gas Mixture}},}\ }in\ \href {\doibase
  10.1007/978-3-319-91548-7_15} {\emph {\bibinfo {booktitle} {Theory, Numerics
  and Applications of Hyperbolic Problems II}}},\ \bibinfo {editor} {edited by\
  \bibinfo {editor} {\bibfnamefont {C.}~\bibnamefont {Klingenberg}}\ and\
  \bibinfo {editor} {\bibfnamefont {M.}~\bibnamefont {Westdickenberg}}}\
  (\bibinfo  {publisher} {Springer International Publishing},\ \bibinfo {year}
  {2018})\ pp.\ \bibinfo {pages} {195--208}\BibitemShut {NoStop}%
\bibitem [{\citenamefont {Groppi}, \citenamefont {Monica},\ and\ \citenamefont
  {Spiga}(2011)}]{Groppi2011}%
  \BibitemOpen
  \bibfield  {author} {\bibinfo {author} {\bibfnamefont {M.}~\bibnamefont
  {Groppi}}, \bibinfo {author} {\bibfnamefont {S.}~\bibnamefont {Monica}}, \
  and\ \bibinfo {author} {\bibfnamefont {G.}~\bibnamefont {Spiga}},\ }\bibfield
   {title} {\enquote {\bibinfo {title} {{A kinetic ellipsoidal BGK model for a
  binary gas mixture}},}\ }\href {\doibase 10.1209/0295-5075/96/64002}
  {\bibfield  {journal} {\bibinfo  {journal} {EPL (Europhysics Letters)}\
  }\textbf {\bibinfo {volume} {96}},\ \bibinfo {pages} {64002} (\bibinfo {year}
  {2011})}\BibitemShut {NoStop}%
\bibitem [{\citenamefont {Brull}(2014)}]{Brull2014}%
  \BibitemOpen
  \bibfield  {author} {\bibinfo {author} {\bibfnamefont {S.}~\bibnamefont
  {Brull}},\ }\bibfield  {title} {\enquote {\bibinfo {title} {{An ellipsoidal
  statistical model for gas mixtures}},}\ }\href {\doibase
  10.4310/cms.2015.v13.n1.a1} {\bibfield  {journal} {\bibinfo  {journal}
  {Communications in Mathematical Sciences}\ }\textbf {\bibinfo {volume}
  {13}},\ \bibinfo {pages} {1--13} (\bibinfo {year} {2014})}\BibitemShut
  {NoStop}%
\bibitem [{\citenamefont {Todorova}\ and\ \citenamefont
  {Steijl}(2019)}]{Todorova2019}%
  \BibitemOpen
  \bibfield  {author} {\bibinfo {author} {\bibfnamefont {B.~N.}\ \bibnamefont
  {Todorova}}\ and\ \bibinfo {author} {\bibfnamefont {R.}~\bibnamefont
  {Steijl}},\ }\bibfield  {title} {\enquote {\bibinfo {title} {{Derivation and
  numerical comparison of Shakhov and Ellipsoidal Statistical kinetic models
  for a monoatomic gas mixture}},}\ }\href {\doibase
  10.1016/j.euromechflu.2019.04.001} {\bibfield  {journal} {\bibinfo  {journal}
  {European Journal of Mechanics, B/Fluids}\ }\textbf {\bibinfo {volume}
  {76}},\ \bibinfo {pages} {390--402} (\bibinfo {year} {2019})}\BibitemShut
  {NoStop}%
\bibitem [{\citenamefont {Wilke}(1950)}]{Wilke1950}%
  \BibitemOpen
  \bibfield  {author} {\bibinfo {author} {\bibfnamefont {C.~R.}\ \bibnamefont
  {Wilke}},\ }\bibfield  {title} {\enquote {\bibinfo {title} {{A Viscosity
  Equation for Gas Mixtures}},}\ }\href {\doibase 10.1063/1.1747673} {\bibfield
   {journal} {\bibinfo  {journal} {The Journal of Chemical Physics 18,}\
  }\textbf {\bibinfo {volume} {18}},\ \bibinfo {pages} {517--519} (\bibinfo
  {year} {1950})}\BibitemShut {NoStop}%
\bibitem [{\citenamefont {Palmer}\ and\ \citenamefont
  {Wright}(2003)}]{Palmer2003}%
  \BibitemOpen
  \bibfield  {author} {\bibinfo {author} {\bibfnamefont {G.~E.}\ \bibnamefont
  {Palmer}}\ and\ \bibinfo {author} {\bibfnamefont {M.~J.}\ \bibnamefont
  {Wright}},\ }\bibfield  {title} {\enquote {\bibinfo {title} {{Comparison of
  Methods to Compute High-Temperature Gas Viscosity Introduction}},}\ }\href
  {\doibase 10.2514/2.6756} {\bibfield  {journal} {\bibinfo  {journal} {Journal
  of Thermophysics and Heat Transfer}\ }\textbf {\bibinfo {volume} {17}}
  (\bibinfo {year} {2003}),\ 10.2514/2.6756}\BibitemShut {NoStop}%
\bibitem [{\citenamefont {Hirschfelder}, \citenamefont {Curtiss},\ and\
  \citenamefont {Bird}(1964)}]{Hirschfelder1964}%
  \BibitemOpen
  \bibfield  {author} {\bibinfo {author} {\bibfnamefont {J.~O.}\ \bibnamefont
  {Hirschfelder}}, \bibinfo {author} {\bibfnamefont {C.~F.}\ \bibnamefont
  {Curtiss}}, \ and\ \bibinfo {author} {\bibfnamefont {R.~B.}\ \bibnamefont
  {Bird}},\ }\href@noop {} {\emph {\bibinfo {title} {{The Molecular Theory of
  Gases and Liquids}}}},\ \bibinfo {edition} {revised ed}\ ed.\ (\bibinfo
  {publisher} {Wiley-Interscience},\ \bibinfo {year} {1964})\ p.\ \bibinfo
  {pages} {1280}\BibitemShut {NoStop}%
\bibitem [{\citenamefont {Kestin}\ \emph {et~al.}(1984)\citenamefont {Kestin},
  \citenamefont {Knierim}, \citenamefont {Mason}, \citenamefont {Najafi},
  \citenamefont {Ro},\ and\ \citenamefont {Waldman}}]{Kestin1984}%
  \BibitemOpen
  \bibfield  {author} {\bibinfo {author} {\bibfnamefont {J.}~\bibnamefont
  {Kestin}}, \bibinfo {author} {\bibfnamefont {K.}~\bibnamefont {Knierim}},
  \bibinfo {author} {\bibfnamefont {E.~A.}\ \bibnamefont {Mason}}, \bibinfo
  {author} {\bibfnamefont {B.}~\bibnamefont {Najafi}}, \bibinfo {author}
  {\bibfnamefont {S.~T.}\ \bibnamefont {Ro}}, \ and\ \bibinfo {author}
  {\bibfnamefont {M.}~\bibnamefont {Waldman}},\ }\href {\doibase
  10.1063/1.555703} {\enquote {\bibinfo {title} {{Equilibrium and Transport
  Properties of the Noble Gases and Their Mixtures at Low Density}},}\ }
  (\bibinfo {year} {1984})\BibitemShut {NoStop}%
\bibitem [{\citenamefont {Capitelli}\ \emph {et~al.}(2000)\citenamefont
  {Capitelli}, \citenamefont {Gorse}, \citenamefont {Longo},\ and\
  \citenamefont {Giordano}}]{Capitelli2000}%
  \BibitemOpen
  \bibfield  {author} {\bibinfo {author} {\bibfnamefont {M.}~\bibnamefont
  {Capitelli}}, \bibinfo {author} {\bibfnamefont {C.}~\bibnamefont {Gorse}},
  \bibinfo {author} {\bibfnamefont {S.}~\bibnamefont {Longo}}, \ and\ \bibinfo
  {author} {\bibfnamefont {D.}~\bibnamefont {Giordano}},\ }\bibfield  {title}
  {\enquote {\bibinfo {title} {{Collision integrals of high-temperature air
  species}},}\ }\href {\doibase 10.2514/2.6517} {\bibfield  {journal} {\bibinfo
   {journal} {Journal of Thermophysics and Heat Transfer}\ }\textbf {\bibinfo
  {volume} {14}},\ \bibinfo {pages} {259--268} (\bibinfo {year}
  {2000})}\BibitemShut {NoStop}%
\bibitem [{\citenamefont {Wright}\ \emph {et~al.}(2005)\citenamefont {Wright},
  \citenamefont {Bose}, \citenamefont {Palmer},\ and\ \citenamefont
  {Levin}}]{Wright2005}%
  \BibitemOpen
  \bibfield  {author} {\bibinfo {author} {\bibfnamefont {M.~J.}\ \bibnamefont
  {Wright}}, \bibinfo {author} {\bibfnamefont {D.}~\bibnamefont {Bose}},
  \bibinfo {author} {\bibfnamefont {G.~E.}\ \bibnamefont {Palmer}}, \ and\
  \bibinfo {author} {\bibfnamefont {E.}~\bibnamefont {Levin}},\ }\bibfield
  {title} {\enquote {\bibinfo {title} {{Recommended Collision Integrals for
  Transport Property Computations Part 1: Air Species}},}\ }\href {\doibase
  10.2514/1.16713} {\bibfield  {journal} {\bibinfo  {journal} {AIAA Journal}\
  }\textbf {\bibinfo {volume} {43}},\ \bibinfo {pages} {2558--2564} (\bibinfo
  {year} {2005})}\BibitemShut {NoStop}%
\bibitem [{\citenamefont {Wright}, \citenamefont {Hwang},\ and\ \citenamefont
  {Schwenke}(2007)}]{Wright2007a}%
  \BibitemOpen
  \bibfield  {author} {\bibinfo {author} {\bibfnamefont {M.~J.}\ \bibnamefont
  {Wright}}, \bibinfo {author} {\bibfnamefont {H.~H.}\ \bibnamefont {Hwang}}, \
  and\ \bibinfo {author} {\bibfnamefont {D.~W.}\ \bibnamefont {Schwenke}},\
  }\bibfield  {title} {\enquote {\bibinfo {title} {{Recommended collision
  integrals for transport property computations part 2: Mars and venus
  entries}},}\ }\href {\doibase 10.2514/1.24523} {\bibfield  {journal}
  {\bibinfo  {journal} {AIAA Journal}\ }\textbf {\bibinfo {volume} {45}},\
  \bibinfo {pages} {281--288} (\bibinfo {year} {2007})}\BibitemShut {NoStop}%
\bibitem [{\citenamefont {Stephani}, \citenamefont {Goldstein},\ and\
  \citenamefont {Varghese}(2012)}]{Stephani2012}%
  \BibitemOpen
  \bibfield  {author} {\bibinfo {author} {\bibfnamefont {K.~A.}\ \bibnamefont
  {Stephani}}, \bibinfo {author} {\bibfnamefont {D.~B.}\ \bibnamefont
  {Goldstein}}, \ and\ \bibinfo {author} {\bibfnamefont {P.~L.}\ \bibnamefont
  {Varghese}},\ }\bibfield  {title} {\enquote {\bibinfo {title} {{Consistent
  treatment of transport properties for five-species air direct simulation
  Monte Carlo/Navier-Stokes applications}},}\ }\href {\doibase
  10.1063/1.4729610} {\bibfield  {journal} {\bibinfo  {journal} {Physics of
  Fluids}\ }\textbf {\bibinfo {volume} {24}},\ \bibinfo {pages} {077101}
  (\bibinfo {year} {2012})}\BibitemShut {NoStop}%
\bibitem [{\citenamefont {Venkattraman}\ and\ \citenamefont
  {Alexeenko}(2012)}]{Venkattraman2012}%
  \BibitemOpen
  \bibfield  {author} {\bibinfo {author} {\bibfnamefont {A.}~\bibnamefont
  {Venkattraman}}\ and\ \bibinfo {author} {\bibfnamefont {A.~A.}\ \bibnamefont
  {Alexeenko}},\ }\bibfield  {title} {\enquote {\bibinfo {title} {{Binary
  scattering model for Lennard-Jones potential: Transport coefficients and
  collision integrals for non-equilibrium gas flow simulations}},}\ }\href
  {\doibase 10.1063/1.3682375} {\bibfield  {journal} {\bibinfo  {journal}
  {Physics of Fluids}\ }\textbf {\bibinfo {volume} {24}} (\bibinfo {year}
  {2012}),\ 10.1063/1.3682375}\BibitemShut {NoStop}%
\bibitem [{\citenamefont {Mathiaud}\ and\ \citenamefont
  {Mieussens}(2016)}]{mathiaud2016fokker}%
  \BibitemOpen
  \bibfield  {author} {\bibinfo {author} {\bibfnamefont {J.}~\bibnamefont
  {Mathiaud}}\ and\ \bibinfo {author} {\bibfnamefont {L.}~\bibnamefont
  {Mieussens}},\ }\bibfield  {title} {\enquote {\bibinfo {title} {A
  fokker--planck model of the boltzmann equation with correct prandtl
  number},}\ }\href@noop {} {\bibfield  {journal} {\bibinfo  {journal} {Journal
  of Statistical Physics}\ }\textbf {\bibinfo {volume} {162}},\ \bibinfo
  {pages} {397--414} (\bibinfo {year} {2016})}\BibitemShut {NoStop}%
\bibitem [{\citenamefont {Burt}\ and\ \citenamefont
  {Boyd}(2006)}]{burt2006evaluation}%
  \BibitemOpen
  \bibfield  {author} {\bibinfo {author} {\bibfnamefont {J.}~\bibnamefont
  {Burt}}\ and\ \bibinfo {author} {\bibfnamefont {I.}~\bibnamefont {Boyd}},\
  }\bibfield  {title} {\enquote {\bibinfo {title} {{Evaluation of a particle
  method for the ellipsoidal statistical Bhatnagar-Gross-Krook equation}},}\
  }in\ \href@noop {} {\emph {\bibinfo {booktitle} {44th AIAA Aerospace Sciences
  Meeting and Exhibit}}}\ (\bibinfo {year} {2006})\ p.\ \bibinfo {pages}
  {989}\BibitemShut {NoStop}%
\bibitem [{\citenamefont {Munz}\ \emph {et~al.}(2014)\citenamefont {Munz},
  \citenamefont {Auweter-Kurtz}, \citenamefont {Fasoulas}, \citenamefont
  {Mirza}, \citenamefont {Ortwein}, \citenamefont {Pfeiffer},\ and\
  \citenamefont {Stindl}}]{Munz2014}%
  \BibitemOpen
  \bibfield  {author} {\bibinfo {author} {\bibfnamefont {C.-D.}\ \bibnamefont
  {Munz}}, \bibinfo {author} {\bibfnamefont {M.}~\bibnamefont {Auweter-Kurtz}},
  \bibinfo {author} {\bibfnamefont {S.}~\bibnamefont {Fasoulas}}, \bibinfo
  {author} {\bibfnamefont {A.}~\bibnamefont {Mirza}}, \bibinfo {author}
  {\bibfnamefont {P.}~\bibnamefont {Ortwein}}, \bibinfo {author} {\bibfnamefont
  {M.}~\bibnamefont {Pfeiffer}}, \ and\ \bibinfo {author} {\bibfnamefont
  {T.}~\bibnamefont {Stindl}},\ }\bibfield  {title} {\enquote {\bibinfo {title}
  {{Coupled Particle-In-Cell and Direct Simulation Monte Carlo method for
  simulating reactive plasma flows}},}\ }\href {\doibase
  10.1016/j.crme.2014.07.005} {\bibfield  {journal} {\bibinfo  {journal}
  {Comptes Rendus M{\'{e}}canique}\ }\textbf {\bibinfo {volume} {342}},\
  \bibinfo {pages} {662--670} (\bibinfo {year} {2014})}\BibitemShut {NoStop}%
\bibitem [{\citenamefont {Fasoulas}\ \emph {et~al.}(2019)\citenamefont
  {Fasoulas}, \citenamefont {Munz}, \citenamefont {Pfeiffer}, \citenamefont
  {Beyer}, \citenamefont {Binder}, \citenamefont {Copplestone}, \citenamefont
  {Mirza}, \citenamefont {Nizenkov}, \citenamefont {Ortwein},\ and\
  \citenamefont {Reschke}}]{fasoulas2019combining}%
  \BibitemOpen
  \bibfield  {author} {\bibinfo {author} {\bibfnamefont {S.}~\bibnamefont
  {Fasoulas}}, \bibinfo {author} {\bibfnamefont {C.-D.}\ \bibnamefont {Munz}},
  \bibinfo {author} {\bibfnamefont {M.}~\bibnamefont {Pfeiffer}}, \bibinfo
  {author} {\bibfnamefont {J.}~\bibnamefont {Beyer}}, \bibinfo {author}
  {\bibfnamefont {T.}~\bibnamefont {Binder}}, \bibinfo {author} {\bibfnamefont
  {S.}~\bibnamefont {Copplestone}}, \bibinfo {author} {\bibfnamefont
  {A.}~\bibnamefont {Mirza}}, \bibinfo {author} {\bibfnamefont
  {P.}~\bibnamefont {Nizenkov}}, \bibinfo {author} {\bibfnamefont
  {P.}~\bibnamefont {Ortwein}}, \ and\ \bibinfo {author} {\bibfnamefont
  {W.}~\bibnamefont {Reschke}},\ }\bibfield  {title} {\enquote {\bibinfo
  {title} {Combining particle-in-cell and direct simulation monte carlo for the
  simulation of reactive plasma flows},}\ }\href@noop {} {\bibfield  {journal}
  {\bibinfo  {journal} {Physics of Fluids}\ }\textbf {\bibinfo {volume} {31}},\
  \bibinfo {pages} {072006} (\bibinfo {year} {2019})}\BibitemShut {NoStop}%
\bibitem [{\citenamefont {Gallis}\ and\ \citenamefont
  {Torczynski}(2011)}]{gallis2011investigation}%
  \BibitemOpen
  \bibfield  {author} {\bibinfo {author} {\bibfnamefont {M.}~\bibnamefont
  {Gallis}}\ and\ \bibinfo {author} {\bibfnamefont {J.}~\bibnamefont
  {Torczynski}},\ }\bibfield  {title} {\enquote {\bibinfo {title}
  {{Investigation of the ellipsoidal-statistical Bhatnagar-Gross-Krook kinetic
  model applied to gas-phase transport of heat and tangential momentum between
  parallel walls}},}\ }\href@noop {} {\bibfield  {journal} {\bibinfo  {journal}
  {Physics of Fluids}\ }\textbf {\bibinfo {volume} {23}},\ \bibinfo {pages}
  {030601} (\bibinfo {year} {2011})}\BibitemShut {NoStop}%
\bibitem [{\citenamefont {Gallis}\ and\ \citenamefont
  {Torczynski}(2000)}]{gallis2000application}%
  \BibitemOpen
  \bibfield  {author} {\bibinfo {author} {\bibfnamefont {M.}~\bibnamefont
  {Gallis}}\ and\ \bibinfo {author} {\bibfnamefont {J.}~\bibnamefont
  {Torczynski}},\ }\bibfield  {title} {\enquote {\bibinfo {title} {{The
  application of the BGK model in particle simulations}},}\ }in\ \href@noop {}
  {\emph {\bibinfo {booktitle} {34th Thermophysics Conference}}}\ (\bibinfo
  {year} {2000})\ p.\ \bibinfo {pages} {2360}\BibitemShut {NoStop}%
\end{thebibliography}%
\end{document}